\documentclass[
  aps,prd,twocolumn,nofootinbib,superscriptaddress,
  longbibliography,floatfix,preprintnumbers
]{revtex4-2}

\usepackage{amsmath,amssymb}
\usepackage{bm}
\usepackage{graphicx}
\usepackage{hyperref}
\usepackage{orcidlink}
\definecolor{linknavy}{rgb}{0.0,0.2,0.55}
\hypersetup{
  colorlinks=true,
  linkcolor=linknavy,
  citecolor=linknavy,
  urlcolor=linknavy,
  filecolor=linknavy,
}
\usepackage{tikz}
\usetikzlibrary{feynmangraphz}
\usetikzlibrary{decorations.pathmorphing,decorations.markings,arrows.meta,calc}
\tikzset{
  midar/.style={postaction={decorate,decoration={markings,
    mark=at position 0.58 with {\arrow{Stealth[length=11pt,width=8pt]}}}}},
  nucl/.style={draw,thick,midar},
  phot/.style={draw,thick,
    decorate,decoration={snake,amplitude=2.2pt,segment length=5pt}},
}
\newcommand{\cutint}[2]{%
  \draw[thick,dashed,dash pattern=on 3pt off 2.5pt] (#1,-#2) -- (#1,#2);}
\newcommand{\dwave}[2]{%
  \draw[thick,double,double distance=1.1pt,decorate,
        decoration={snake,amplitude=1.4pt,segment length=6pt}] (#1)--(#2);%
  \draw[draw=none,midar] (#1)--(#2);%
}

\newcommand{\npdg}{n+p\to d+\gamma}
\newcommand{\dpHe}{d(p,\gamma){}^{3}\mathrm{He}}
\newcommand{\prel}{p_{\mathrm{rel}}}
\newcommand{\Egam}{E_{\gamma}}
\newcommand{\muN}{\mu_{N}}
\newcommand{\kappaV}{\kappa_{V}}
\newcommand{\Mone}{M1}
\newcommand{\Eone}{E1}
\newcommand{\Etwo}{E2}
\newcommand{\Mtwo}{M2}
\newcommand{\Ethree}{E3}
\newcommand{\Mthree}{M3}
\newcommand{\ad}[2]{\langle #1 #2\rangle}
\newcommand{\sq}[2]{[#1 #2]}
\newcommand{\sigth}{\sigma_{\mathrm{th}}}

\newcommand{\AS}{A_{S}}
\newcommand{\CSv}{\tilde{C}_{S}}
\newcommand{\CDv}{\tilde{C}_{D}}
\newcommand{\Cone}{\tilde{C}_{1}}
\newcommand{\Ctwo}{\tilde{C}_{2}}
\newcommand{\pstar}{p_{\ast}}

\begin{document}

\preprint{UCI-HEP-TR-2026-10}

\title{Deuterium Production in an Effective Field Theory\\ Constructed from On-Shell Amplitudes}

\author{Tim M.P.~Tait\,\orcidlink{0000-0003-3002-6909}}
\email{ttait@uci.edu}
\affiliation{Department of Physics and Astronomy,
University of California, Irvine, CA 92697, USA}

\date{\today}

\begin{abstract}
We compute the deuterium-production reaction $\npdg$ in an effective field
theory whose degrees of freedom are the nuclear states themselves: the
amplitude is assembled from on-shell three-point vertices, glued across its
factorization channels, and completed by the contact terms consistent with the
symmetries. The deuteron enters through the $d$--$n$--$p$ vertex, normalized to
the measured asymptotic normalization coefficient.
Rescattering of the nucleon pair is resummed dispersively, leaving two
short-distance constants, an isovector magnetic and an electric dipole
contact interaction. A joint Bayesian fit to the thermal capture measurements and the
SLEGS photodisintegration data finds both of natural size and determines the
thermonuclear rate to $0.22$--$0.24\%$ across the nucleosynthesis window,
including systematics spanning the defensible treatments of the SLEGS data and of
the $P$-wave rescattering. Truncating the expansion is bounded separately at
$0.12\%$, of which the next order of contact terms---degenerate with the two fitted
constants---supplies $0.03\%$, for a total theory uncertainty of $0.25$--$0.27\%$.
Propagated through a BBN network, the rate shifts the
predicted primordial deuterium by $-0.06\%$ and cuts this reaction's
contribution to the $\mathrm{D/H}$ uncertainty from $0.089\%$ to $0.050\%$,
retiring it from the primordial $\mathrm{D/H}$ error budget for practical purposes.
\end{abstract}

\maketitle

\section{Introduction}

Big Bang nucleosynthesis (BBN) assembled the lightest nuclei during the first
minutes of cosmic history~\cite{Alpher:1948ve,Wagoner:1966pv}, and the abundances it
left behind test the standard cosmology with percent-level
precision~\cite{Cyburt:2015mya,Pitrou:2018cgg}. Primordial deuterium carries some of the sharpest
information, with absorption spectra of metal-poor systems along quasar sightlines
giving $10^{5}\,\mathrm{D/H}=2.527\pm0.030$~\cite{Cooke:2017cwo}. The predicted
abundance falls steeply with the baryon density $\omega_b\equiv\Omega_bh^2$, so
primordial deuterium returns a measurement of $\omega_b$ that owes nothing to
the cosmic microwave background~\cite{Planck:2018vyg}. Confronting the BBN
prediction with the observed abundance therefore closes a loop, and their agreement is an important test of
$\Lambda$CDM back to $\sim$~MeV temperatures. Disagreement would be a clue pointing to physics that alters the expansion rate
or particle content/interactions during the nucleosynthesis epoch~\cite{Pospelov:2010hj}.

On the theory side, the limiting inputs are the rates of nuclear reactions. A
single reaction, radiative neutron capture $\npdg$, produces deuterium, whereas three
reactions burn it: $d(p,\gamma){}^3\mathrm{He}$, $d(d,n){}^3\mathrm{He}$, and
$d(d,p)t$~\cite{Cyburt:2015mya,Yeh:2020mgl}. Every heavier nucleus starts from
deuterium, so the capture process opens nucleosynthesis itself. The burning rates
currently dominate the theoretical error budget on
$\mathrm{D/H}$~\cite{Mossa:2020gjc,Pitrou:2020etk,Pisanti:2020efz,Launders:2026ciu}.
Holding the production rate to few-per-mille precision keeps it subdominant as
the burning data improve. The recent high-precision measurement of deuteron
photodisintegration at the Shanghai Laser Electron Gamma Source
(SLEGS)~\cite{Chen:2025oyj}, analyzed by the collaboration in dibaryon
EFT~\cite{Ando:2005cz}, makes that precision attainable: fit jointly with the
thermal capture cross section, it pins the electric amplitude across the upper
nucleosynthesis window.

Nucleosynthesis operates at $E_{\mathrm{cm}}\sim0.01$--$1\,\mathrm{MeV}$, where
the relative momentum sits far below the pion mass. Pions are thus inert
degrees of freedom, and contact interactions among nucleons capture the residual strong
dynamics~\cite{Chen:1999tn}. Neutron capture is among the most completely
studied processes in nuclear theory. In fact, it is the first nuclear reaction computed directly from lattice QCD~\cite{Beane:2015yha}.
Effective-range theory reproduces the
thermal cross section at the ten-percent level~\cite{Bethe:1950jm}. The missing
ten percent is the classic meson-exchange two-body current~\cite{Riska:1972zz},
as derived in chiral effective Lagrangians~\cite{Park:1994sr}. Pionless
effective field theory organizes both pieces into a controlled expansion and
delivers the cross section to the one-percent level across the BBN
window~\cite{Chen:1999tn,Rupak:1999rk}. Its dibaryon variant supplies the rate
parametrizations used by the network codes~\cite{Ando:2005cz}. Chiral effective
field theory~\cite{Epelbaum:2008ga,Machleidt:2011zz} reproduces the same
observables with quantified truncation
uncertainties~\cite{Acharya:2021lrv,Du:2022zds}. Conventional {\it ab initio}
calculations with phenomenological interactions and exchange currents describe
the full set of $A=2$ and $3$ electromagnetic observables~\cite{Marcucci:2005zc}.
The same machinery, with consistent one- and two-body currents, now reaches
electromagnetic radii up to $A=10$~\cite{King:2025akz} and electroweak response
functions in ${}^{16}\mathrm{O}$ and
${}^{40}\mathrm{Ca}$~\cite{Acharya:2024xah,Sobczyk:2021dwm}; in
${}^{12}\mathrm{C}$ the two-body terms carry roughly a third of the neutral-weak
strength~\cite{Lovato:2014eva}.

The same theoretical maturity extends across the neighboring light-nucleus
captures of primordial and stellar
interest~\cite{Adelberger:2010qa,Acharya:2024lke}. For the burning reaction
$\dpHe$, the underground LUNA program covers the solar Gamow peak and most of
the BBN window~\cite{Casella:2002yej,Mossa:2020gjc,Stockel:2024hde}, with
several groups extending the reach to higher
energies~\cite{Tisma:2019acf,Turkat:2021qmq,Schmid:1997zz}. The
hyperspherical-harmonics {\it ab initio} calculation~\cite{Marcucci:2015yla}
lands systematically above the post-LUNA data, and BBN network codes disagree on
which to trust~\cite{Pitrou:2018cgg,Gariazzo:2021iiu,Moscoso:2021xog}. Chiral-EFT reaction
theory reaches nucleon--deuteron capture~\cite{Skibinski:2006gy} and, more
recently, $d(\alpha,\gamma){}^{6}\mathrm{Li}$~\cite{Hebborn:2022iiz} and
${}^{3}\mathrm{He}(\alpha,\gamma){}^{7}\mathrm{Be}$~\cite{Atkinson:2024zrm}.
Halo and cluster EFTs treat radiative captures involving heavier
clusters~\cite{Bertulani:2002sz,Bedaque:2003wa,Rupak:2016mmz}. Any framework
for nuclear reactions must confront this body of work quantitatively. With its
sub-percent thermal anchor and its modern photodisintegration data, $\npdg$ is the sharpest comparison.

In Ref.~\cite{Tait:2026guk}, we developed an effective field theory for the radiative
capture $\dpHe$ in which the amplitude is assembled with modern on-shell
methods, treating the individual states of the nuclei themselves as the degrees of freedom.  Rather than writing down a Lagrangian and deriving a set of Feynman
rules from it, the amplitude approach enumerates the on-shell three-point vertices consistent with the
quantum numbers of each nucleus. These glue into the four-point amplitude on its
factorization channels, and local contact terms allowed by gauge invariance and the
symmetries complete the result, encoding deviations from the approximation that nuclei are point-like at the wavelengths of interest. The low-energy constants are the coefficients of
those structures, matched to data. Here we apply that framework to $\npdg$. The
calculation is an efficient test of the method: almost every ingredient is
already in hand, and the one genuinely new vertex (the neutron's magnetic photon coupling) has a transparent physical structure.

In some ways, $\npdg$ is a simpler process than $\dpHe$. First, the neutron is neutral,
so the entrance channel carries no Coulomb interaction -- there is no Sommerfeld
factor and no penetration barrier.
Second, the dominant near-threshold transition is the isovector magnetic dipole
$\Mone$, fixed by a single precisely measured quantity, the thermal capture cross
section. Third, the strong vertex that carries the deuteron is the same
$d$--$n$--$p$ amplitude that fixes the deuteron leg of the $d+d$ reactions; its
normalization is matched to the deuteron asymptotic normalization coefficient (ANC). 

This paper is organized as follows. Section~\ref{sec:framework} constructs the
amplitude: the three-point vertices, the factorized four-point amplitude with
its Ward identity, and the contact currents. Section~\ref{sec:xsec} assembles
these into the cross section, dressed by the initial-state interactions.
Section~\ref{sec:results} confronts it with data: a joint Bayesian fit to the
thermal capture measurements and the SLEGS photodisintegration data determines
the magnetic and electric contact coefficients, and delivers the cross section, the
astrophysical $S$-factor $S(E)$, and the thermonuclear reaction rate.
Section~\ref{sec:discussion} provides a summary, discussion,
and outlines natural extensions.  Details concerning the resummation of the initial
state (re)scattering are in the Appendix.

\section{EFT and Amplitudes}
\label{sec:framework}

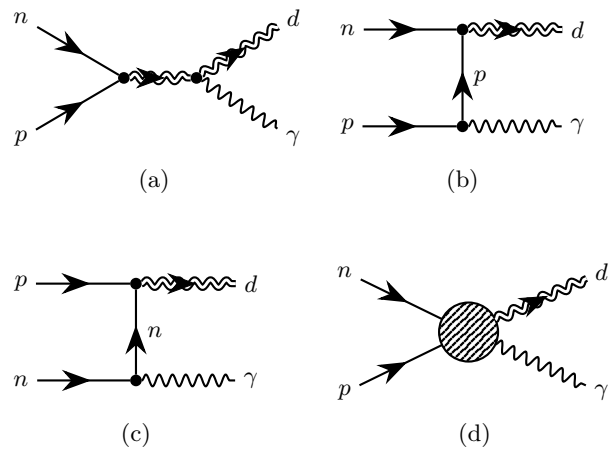
\begin{figure}[t]
\centering
\begin{tikzpicture}[fmg, every node/.append style={font=\small}, scale=0.8]
\begin{scope}[local bounding box=A]
  \path (-0.3,1.0) node (n) {$n$} (-0.3,-1.0) node (p) {$p$}
        (1.4,0) node[vertex,name=S] {} (2.6,0) node[vertex,name=E] {}
        (4.2,1.0) node (d) {$d$} (4.2,-1.0) node (g) {$\gamma$};
  \draw[nucl] (n) -- (S); \draw[nucl] (p) -- (S);
  \dwave{S}{E} \dwave{E}{d} \draw[phot] (E) -- (g);
  \node at (1.9,-1.7) {(a)};
\end{scope}
\begin{scope}[shift={(5.4,0)},local bounding box=B]
  \path (-0.3,0.8) node (n) {$n$} (-0.3,-0.8) node (p) {$p$}
        (1.6,0.8) node[vertex,name=S] {} (1.6,-0.8) node[vertex,name=E] {}
        (3.5,0.8) node (d) {$d$} (3.5,-0.8) node (g) {$\gamma$};
  \draw[nucl] (n) -- (S); \draw[nucl] (p) -- (E); \draw[nucl] (E) -- (S);
  \dwave{S}{d} \draw[phot] (E) -- (g);
  \node[right=1pt] at (1.6,0) {$p$}; \node at (1.6,-1.7) {(b)};
\end{scope}
\begin{scope}[shift={(0,-4.2)},local bounding box=C]
  \path (-0.3,0.8) node (p) {$p$} (-0.3,-0.8) node (n) {$n$}
        (1.6,0.8) node[vertex,name=S] {} (1.6,-0.8) node[vertex,name=E] {}
        (3.5,0.8) node (d) {$d$} (3.5,-0.8) node (g) {$\gamma$};
  \draw[nucl] (p) -- (S); \draw[nucl] (n) -- (E); \draw[nucl] (E) -- (S);
  \dwave{S}{d} \draw[phot] (E) -- (g);
  \node[right=1pt] at (1.6,0) {$n$}; \node at (1.6,-1.7) {(c)};
\end{scope}
\begin{scope}[shift={(5.4,-4.2)},local bounding box=D]
  \path (-0.35,1.0) node (n) {$n$} (-0.35,-1.0) node (p) {$p$}
        (1.7,0) node[blob,minimum size=2.4em,name=Bl] {}
        (3.9,1.0) node (d) {$d$} (3.9,-1.0) node (g) {$\gamma$};
  \draw[nucl] (n) -- (Bl); \draw[nucl] (p) -- (Bl);
  \dwave{Bl}{d} \draw[phot] (Bl) -- (g);
  \node at (1.8,-1.7) {(d)};
\end{scope}
\end{tikzpicture}
\caption{Diagrammatic summary of contributions to $\npdg$. Diagrams (a)--(c) represent the factorized channels: radiation off the (a) deuteron, (b) proton, and (c) neutron.
Diagram (d) is the contact (boundary) term. The deuteron is indicated by a double wavy line, nucleons by single lines, and the photon by a single wavy line.}
\label{fig:diagrams}
\end{figure}

The amplitude for $\npdg$ is the sum of two pieces: the factorized piece glues
on-shell three-point vertices across the channels on which the four-point
amplitude develops a pole, whereas the boundary piece collects local contact terms.
These are summarized diagrammatically in Fig.~\ref{fig:diagrams}.
The construction parallels the corresponding one for $\dpHe$ in
Ref.~\cite{Tait:2026guk}, and inherits some of its derivations.

We use the massive spinor-helicity conventions of
Ref.~\cite{Arkani-Hamed:2017jhn}: bold spinors
$|\bm{i}^{I}\rangle,|\bm{i}^{I}]$ carry the (implicitly symmetrized) $SU(2)$ little-group
indices of the massive legs, and amplitudes are polynomials in $\langle\,\rangle,[\,]$. All
legs are defined as incoming, so the capture process has $n$ and $p$ as particles and the deuteron
entering as $\bar d$.

\subsection{Three-Point Vertices}
\label{sec:vertices}

\subsubsection{Strong Vertex} 
\label{sec:strongvertex}

The strong vertex is the $d$--$n$--$p$ coupling, an
all-massive three-point amplitude of spins $(1,\tfrac12,\tfrac12)$. It reduces
in the same way as the $d$--$p$--${}^3\mathrm{He}$ vertex of Ref.~\cite{Tait:2026guk}.
The naive count $(2S_{d}{+}1)(2S_{n}{+}1)(2S_{p}{+}1)=12$ collapses on the
three-particle locus to four independent monomials,
$\ad{\bm{d}}{\bm{n}}\ad{\bm{d}}{\bm{p}}$, $\sq{\bm{d}}{\bm{n}}\sq{\bm{d}}{\bm{p}}$,
$\ad{\bm{d}}{\bm{n}}\sq{\bm{d}}{\bm{p}}$ and $\sq{\bm{d}}{\bm{n}}\ad{\bm{d}}{\bm{p}}$.
All three legs carry positive parity, so parity conservation keeps the two
bracket-swap-even combinations and discards the two odd ones:
\begin{equation}
\begin{split}
  \mathcal{E}^{(1)} &= \ad{\bm{d}}{\bm{n}}\ad{\bm{d}}{\bm{p}} + \sq{\bm{d}}{\bm{n}}\sq{\bm{d}}{\bm{p}},\\
  \mathcal{E}^{(2)} &= \ad{\bm{d}}{\bm{n}}\sq{\bm{d}}{\bm{p}} + \sq{\bm{d}}{\bm{n}}\ad{\bm{d}}{\bm{p}},
\end{split}
\end{equation}
the deuteron little-group indices symmetrized, with
\begin{equation}
  M_3=\big(\Cone\,\mathcal{E}^{(1)}+\Ctwo\,\mathcal{E}^{(2)}\big)/m_{d},
  \label{eq:strongbasis}
\end{equation}
normalized to the inverse deuteron mass $1/m_{d}$, so that the couplings themselves are dimensionless. Diagonalizing the
$n$+$p$ partial-wave content,
\begin{equation}
\begin{aligned}
  \mathcal{E}^{(S)} &= \mathcal{E}^{(1)}{-}\mathcal{E}^{(2)}
    = \big(\ad{\bm{d}}{\bm{n}}{-}\sq{\bm{d}}{\bm{n}}\big)
      \big(\ad{\bm{d}}{\bm{p}}{-}\sq{\bm{d}}{\bm{p}}\big),\\
  \mathcal{E}^{(D)} &= \mathcal{E}^{(1)}{+}\mathcal{E}^{(2)}
    = \big(\ad{\bm{d}}{\bm{n}}{+}\sq{\bm{d}}{\bm{n}}\big)
      \big(\ad{\bm{d}}{\bm{p}}{+}\sq{\bm{d}}{\bm{p}}\big),
\end{aligned}
  \label{eq:strongSD}
\end{equation}
gives the $(L,s)=(0,1)$ ${}^{3}S_{1}$ and $(2,1)$ ${}^{3}D_{1}$ states of the
$n$+$p$ system---the deuteron's $S$ and $D$ states---with couplings
$\CSv=\tfrac12(\Cone-\Ctwo)$ and $\CDv=\tfrac12(\Cone+\Ctwo)$.

The couplings $\CSv,\CDv$ set the residue of the $n$--$p$ elastic amplitude at the deuteron pole, which can be matched to the ANC:
\begin{equation}
  |M_3(d\!\leftrightarrow\! np)|^2_{\mathrm{phys}}
  = 8\pi\,\frac{m_d\,m_n\,m_p}{\mu^2}\,\AS^2 ,
  \label{eq:residue}
\end{equation}
with $\mu=m_n m_p/(m_n+m_p)$ the reduced mass and
$\AS=0.8846\,\mathrm{fm}^{-1/2}$ the measured $^3S_1$ ANC~\cite{deSwart:1995ui}.  
As a check, Eq.~\eqref{eq:residue} carried through the assembly of Sec.~\ref{sec:xsec} reproduces the textbook Bethe--Peierls~\cite{Bethe:1935diplon} photodisintegration cross section exactly in the zero-range limit.

The $d$-wave coupling follows from the asymptotic $D/S$ ratio $\CDv/\CSv=\eta_d =0.02534$~\cite{deSwart:1995ui}, also adopted by Ref.~\cite{Rupak:1999rk}. There is also a direct
measurement~\cite{Rodning:1990zz} resulting in $\eta_{d}=0.0256(4)$, about $1\%$ higher.
The difference is immaterial for our purposes.

\subsubsection{Electromagnetic Interactions} 

The proton and deuteron electromagnetic vertices are derived in Ref.~\cite{Tait:2026guk}. The photon
couples to each leg through a vertex with two equal-mass lines $\bm{a},\bm{b}$
and the photon $\gamma$. For a spin-$\tfrac12$ nucleon $X$ each photon helicity
carries two structures,
\begin{eqnarray}
  M^{+1}_{X,\,IJ} &=& Z_{X}e\,x\,\ad{\bm{a}^{I}}{\bm{b}^{J}}
     + \frac{e\,\kappa_{X}}{2m_{X}}\,\sq{\bm{a}^{I}}{\gamma}\sq{\bm{b}^{J}}{\gamma},
     \nonumber\\
  M^{-1}_{X,\,IJ} &=& Z_{X}e\,\tilde{x}\,\sq{\bm{a}^{I}}{\bm{b}^{J}}
     + \frac{e\,\kappa_{X}}{2m_{X}}\,\ad{\bm{a}^{I}}{\gamma}\ad{\bm{b}^{J}}{\gamma},
  \label{eq:photonN}
\end{eqnarray}
where $x$ is the standard massless-leg factor and $\tilde{x}=-1/x$ its conjugate
(the per-channel factors appear in Sec.~\ref{sec:fact}). The charge $Z_{X}$ is fixed
by the Ward identity, whereas the dimensionless anomalous moment $\kappa_{X}$ is a free
low energy constant (LEC). The physical quantity entering the magnetic-dipole transition is the total
moment,
\begin{equation}
  \mu_{X} = \big(Z_{X}+\kappa_{X}\big)\frac{m_{p}}{m_{X}}\quad[\muN],
  \qquad \muN=\frac{e}{2m_{p}},
  \label{eq:totalmoment}
\end{equation}
so $\mu_{p}=1+\kappa_{p}=2.793$ with $\kappa_{p}=1.793$ for the proton, and $\kappa_n=\mu_n=-1.913$ is the full neutron moment.

The deuteron, a
charged spin-$1$ field, is described by three form factors (little-group indices
suppressed),
\begin{eqnarray}
  M^{+1}_{d} &=& \frac{Z_{d}e}{m_{d}}\,x\,\ad{\bm{a}}{\bm{b}}^{2}
         + \frac{e\,g_{1}^{d}}{m_{d}^{2}}\,x^{2}\,\ad{\gamma}{\bm{a}}\ad{\gamma}{\bm{b}}\ad{\bm{a}}{\bm{b}}
         \nonumber\\
         && +\, \frac{e\,g_{2}^{d}}{m_{d}^{3}}\,x^{3}\,\ad{\gamma}{\bm{a}}^{2}\ad{\gamma}{\bm{b}}^{2},
  \label{eq:photond}
\end{eqnarray}
with charge $Z_{d}=+1$. The dimensionless magnetic dipole $g_{1}^{d}$ is fixed
by the deuteron moment $\mu_{d}=0.857$. The dimensionless electric quadrupole
$g_{2}^{d}$ is fixed by the deuteron quadrupole moment
$\mathcal{Q}_{d}=+0.286\,e\,\mathrm{fm}^{2}$.

\subsection{Factorized Amplitude and Ward Identity}
\label{sec:fact}

On-shell factorization reconstructs the factorizable part of the amplitude from three poles, one per leg
radiating, as shown schematically in Fig.~\ref{fig:diagrams}. Each glues the
strong vertex to a photon vertex across the internal line. The proton and
deuteron channels carry massless-leg factors, written in terms of a reference spinor $\eta$,
\begin{equation}
  x_p=\frac{\langle\eta|p|\gamma]}{m_p\,\ad{\eta}{\gamma}},\qquad
  x_d=\frac{\langle\eta|\bar d|\gamma]}{m_d\,\ad{\eta}{\gamma}},
  \qquad \langle\eta|i|\gamma]\equiv\ad{\eta}{\bm i}\sq{\bm i}{\gamma},
  \label{eq:xfact}
\end{equation}
where the massive little-group indices on each leg $i$ are contracted. The dependence on the arbitrary spinor
$\eta$ is the manifestation of the freedom to choose a QED gauge.

In spinor variables, a massless spin-1 leg with momentum $k_{\alpha\dot\alpha}=\lambda_{\alpha}\tilde\lambda_{\dot\alpha}$ carries the
positive-helicity polarization
\begin{equation}
  e^{+}_{\alpha\dot\alpha}
  = \frac{\eta_{\alpha}\,\tilde\lambda_{\dot\alpha}}{\ad{\eta}{\lambda}},
  \label{eq:polref}
\end{equation}
with $\eta$ the arbitrary reference spinor.  Shifting $\eta$ along the momentum, $\eta\to\eta+b\lambda$, results in
\begin{equation}
  e^{+}\;\longrightarrow\;e^{+}
  +\frac{b}{\ad{\eta}{\lambda}}\;k_{\alpha\dot\alpha}.
  \label{eq:gaugeshift}
\end{equation}
This is the photon's gauge transformation $\varepsilon\to\varepsilon+\alpha k$ written in
spinors: one free component of $\eta$ corresponding to the gauge parameter. 

Under the gauge transformation Eq.~\eqref{eq:gaugeshift} the factorized amplitude varies by
$\delta M_{\mathrm{fact}}=(b/\ad{\eta}{\lambda})\,k_\mu M^{\mu}_{\mathrm{fact}}$,
with
\begin{equation}
  k_\mu M^\mu_{\mathrm{fact}}
  = e\sum_{i=\{p,\bar d\}} Q_{i}\Big[V^{(i)}_{\mathrm{strong}}
                                      -V^{(0)}_{\mathrm{strong}}\Big],
  \label{eq:ward}
\end{equation}
where $V^{(0)}_{\mathrm{strong}}$ is the vertex $V_{\mathrm{strong}}=(\Cone\mathcal{E}^{(1)}+\Ctwo\mathcal{E}^{(2)})/m_{d}$
at unshifted kinematics and
$V^{(i)}_{\mathrm{strong}}$ the same vertex at the point shifted by the photon momentum
when leg $i$ radiates. 
The unshifted piece cancels by charge conservation,
$\sum_{i}Q_{i}=Q_{p}+Q_{\bar d}=1-1=0$, but the shifted piece does not cancel, because
$V_{\mathrm{strong}}$ is momentum dependent. The boundary term $\mathcal{B}_W$ promotes $\partial$ in the $\Cone$ vertex
to the covariant derivative $\partial\to\partial-ieQA$ on the charged legs.  It contains its own $\eta$ dependence, arranged so that 
$\delta(M_{\mathrm{fact}}+\mathcal{B}_W)=0$.  The $\Ctwo$ term, being derivative-free, is already gauge invariant. 
$\mathcal{B}_W$ has the closed bracket form:
\begin{widetext}
\begin{equation}
  \mathcal{B}_{W}^{+1} = -\,\frac{e Z_{d}\,\Cone}{2\,m_{d}^{2}\,\ad{\eta}{\gamma}}\,
  \mathrm{sym}_{(I_{1}I_{2})}\Big\{
  \sq{\bm{d}^{I_{2}}}{\gamma}\big(\ad{\bm{n}}{\bm{d}^{I_{1}}}\ad{\eta}{\bm{p}}
  +\ad{\bm{n}}{\eta}\ad{\bm{d}^{I_{1}}}{\bm{p}}\big)
  -\ad{\bm{d}^{I_{1}}}{\eta}\big(\sq{\bm{n}}{\bm{d}^{I_{2}}}\sq{\gamma}{\bm{p}}
  +\sq{\bm{n}}{\gamma}\sq{\bm{d}^{I_{2}}}{\bm{p}}\big)\Big\},
  \label{eq:BWnpdg}
\end{equation}
\end{widetext}
with the negative-helicity form obtained by taking
$ \ad{\,}{}\!\leftrightarrow\!\sq{\,}{}$ and $\eta\to\tilde\eta$.
Including $\mathcal{B}_W$ renders the amplitude
$\eta$-independent component by component---equivalent to imposing the Ward identity, $k_\mu\mathcal{M}^\mu=0$.\footnote{$\mathcal{B}_{W}$
does not touch the magnetic sector. Projected on the entrance channels it is
entirely $^{3}S_{1}$ ($T=0$) at threshold---the isoscalar sector its
couplings ($Q$, $\mu_{d}$, $Q_{d}$) select---with a $^{1}S_{0}$ component at
the $10^{-14}$ level of its norm, pure recoil. It therefore enters neither
the isovector $\Mone$ of Eq.~\eqref{eq:UM1np} nor the rescattering dressing
of Sec.~\ref{sec:isi}.}

A two-nucleon
state must be totally antisymmetric, so its orbital, spin and isospin labels obey
$(-1)^{L+S+T}=-1$. At the energies of interest the $n + p$ pair dominantly enters in $s$-wave,
in two channels: the spin singlet $^{1}S_{0}$, which must carry $T=1$, and the spin
triplet $^{3}S_{1}$, which must carry $T=0$. The deuteron is the $^{3}S_{1}$ bound
state and is therefore isoscalar. $\Mone$ photon emission flips the nucleon spins while
leaving the orbital motion unchanged, so it connects $^{1}S_{0}\!\to\!{}^{3}S_{1}$;
because the initial state has $T=1$ and the final state $T=0$, that transition must
change isospin by one unit. 
In the factorized amplitude, the photon attaches to the magnetic moment of a single nucleon, and that one-body $\Mone$
operator splits into an isoscalar piece proportional to $\mu_{p}+\mu_{n}=0.88\,\muN$ and an isovector piece proportional to $\mu_{p}-\mu_{n}=4.71\,\muN$; only the isovector
piece supplies the required $\Delta T=1$. Its isoscalar partner is suppressed twice
over: the net moment is five times smaller, and to leading order the isoscalar operator is
proportional to the total spin, which commutes with the Hamiltonian and so cannot
connect the $^{3}S_{1}$ scattering state to the $^{3}S_{1}$ bound state.

\subsection{Boundary Terms}
\label{sec:bdy}

The EFT is an expansion in the finite size of the nuclei, which in this case is characterized by the deuteron breakup momentum~\cite{Tait:2026guk}
\begin{equation}
  \pstar=\sqrt{2\mu B}=\gamma=45.7\,\mathrm{MeV},
  \label{eq:pstar}
\end{equation}
with $\mu=m_{n}m_{p}/(m_{n}+m_{p})=469.5\,\mathrm{MeV}$ the incoming reduced mass. It coincides with the deuteron binding momentum $\gamma$ because the
incoming particles are the deuteron's own constituents. Throughout we write
$m_{N}\equiv2\mu=938.92\,\mathrm{MeV}$ for the isospin-averaged nucleon mass---the
$n$--$p$ mass splitting is well below the accuracy to which we work. Corrections occur at
one-loop at this scale, leading to an effective cutoff of order $\sqrt{4\pi}\,\pstar$.
We normalize higher order EFT corrections by the appropriate power of
\begin{equation}
  \Lambda \;=\; \sqrt{4\pi}\,\pstar \;=\; 162.0\,\mathrm{MeV}
  \label{eq:Lambda}
\end{equation}
times a dimensionless Wilson coefficient. The fact that $\Lambda$ lands near the pion mass
$m_{\pi}\approx140\,\mathrm{MeV}$, the breakdown scale of pionless EFT, is the familiar statement that the deuteron is a natural shallow bound state. 

Fig.~\ref{fig:diagrams}(d) represents both $\mathcal{B}_W$ and also higher order EFT corrections encoding the internal structure of the nuclear states and their short
distance interactions, including free LECs for electromagnetic two-body corrections.
We enumerate their amplitude structure at a given power of $1/ \Lambda$ with the helicity-category algorithm of
Ref.~\cite{DeAngelis:2022qco}.  At dimension six, the eight
non-vanishing helicity categories live in ten structures; invariance under
parity pairs them into five independent currents. With legs
$(1,2,3,4)=(\bar d,n,p,\gamma)$,
\begin{align}
  \mathcal{B}_{1} &= \ad{\bm{1}}{4}^{2}\ad{\bm{2}}{\bm{3}}, &
  \mathcal{B}_{2} &= \ad{\bm{1}}{\bm{2}}\ad{\bm{1}}{4}\ad{\bm{3}}{4}, \nonumber\\
  \mathcal{B}_{3} &= \ad{\bm{1}}{4}^{2}\sq{\bm{2}}{\bm{3}}, &
  \mathcal{B}_{4} &= \ad{\bm{1}}{4}\ad{\bm{2}}{4}\sq{\bm{1}}{\bm{3}}, \nonumber\\
  \mathcal{B}_{5} &= \ad{\bm{1}}{4}\ad{\bm{3}}{4}\sq{\bm{1}}{\bm{2}}, &&
  \label{eq:dw7basis}
\end{align}
each entering the amplitude as $(e\,c_{a} /\Lambda^{3})\,\mathcal{B}_{a}$, with strength parameterized by the dimensionless coefficients $c_{1},\dots,c_{5}$.

To better connect with the selection rules, these boundary terms are reorganized based on their angular momentum and nuclear isospin.
The $^{1}S_{0}$ entrance demands the $\Delta T=1$ combination---the isovector two-body
$\Mone$ current, the $L_{1}$ of pionless EFT~\cite{Chen:1999tn}---while the $^{3}S_{1}$ entrance supports an isoscalar operator of its own. Projecting the structures of
Eq.~\eqref{eq:dw7basis} onto two-nucleon channels identifies the combination that
carries each current,
\begin{align}
  c^{V}_{\Mone} &\propto 2c_{1}-c_{2}+2c_{3}+c_{4}-c_{5}
    \qquad (^{1}S_{0},\;T=1), \label{eq:cM1V}\\
  c_{\Etwo} &\propto \phantom{2c_{1}-{}}c_{2}\phantom{{}+2c_{3}}+c_{4}+c_{5}
    \qquad (^{3}S_{1},\;T=0). \label{eq:cE2}
\end{align}
That isoscalar operator is an $\Etwo$, and there is no isoscalar $\Mone$ contact to
go with it: a Jacob--Wick projection of Eq.~\eqref{eq:dw7basis} finds $\Mone$
strength in the $^{3}S_{1}$ channel only at relative order $\Egam/m_{N}$. The same
orthogonality that suppresses the one-body isoscalar operator above leaves no
short-distance partner behind. The $p$-wave
entrance takes the rest: $2c_{1}-c_{2}-2c_{3}$ in $^{3}P_{0}$, carrying $\Eone$ alone;
$c_{4}-c_{5}$ in $^{3}P_{1}$, carrying $\Mtwo$ alone to the same accuracy; and
$c_{2}$ in $^{1}P_{1}$, likewise $\Mtwo$. The five coefficients thus map one to one
onto five multipole channels, with none left over, and the electric dipole reaches a
single entrance channel---a point we return to in Sec.~\ref{sec:isi}.
All three absences have one origin: the photon reaches the deuteron through
$\ad{\bm{1}}{4}$, which annihilates one of its two little-group indices exactly.
The deuteron polarization is then never free, and the surviving $J=1$ amplitudes
are locked onto the quadrupole; the dipole enters only at
$\mathcal{O}(\Egam/m_{N})$.
The near-threshold unpolarized cross section depends on the single combination $c^{V}_{\Mone}$
and, through the $\Eone$, on $2c_{1}-c_{2}-2c_{3}$.
The $\Etwo$ of Eq.~\eqref{eq:cE2} and the remaining $p$-wave
combinations would contribute to observables sensitive to multipole interference---angular
distributions and polarization asymmetries.

At the next order, the leading structures dressed by one momentum insertion
$\langle i|p_{j}|k]$ span a twelve-dimensional space per photon helicity. Five
of its directions reduce on shell to the leading structures times a mass and
only renormalize $c_{1},\dots,c_{5}$; the remaining seven are new boundary
terms of the form $(e\,\tilde{c}_{a}/\Lambda^{4})\,\widetilde{\mathcal{B}}_{a}$.
Near threshold the insertion's scalar part
reduces to $E_{\gamma}$ times a leading structure, so these operators supply
the contact currents' first energy dependence---for the isovector $\Mone$, the
slope the dibaryon theory calls $\tilde{L}_{np}$~\cite{Rupak:1999rk}. The
$\Egam^{2}$ growth of the leading $\Eone$ contact in the cross section
(Sec.~\ref{sec:isi}) is, by contrast, leading-order kinematics.  A $^{3}D_{1}$ admixture to
$^{3}S_{1}$ enters the $s$-wave multipole, but is suppressed by ${\cal O}(10^{-6})$. 
Two $p_{\rm rel}$-suppressed channels also appear---$^{3}P_{2}$, carrying $\Eone$ and
$\Mtwo$ together with the first $\Ethree$,
and the $d$-wave entrance $^{1}D_{2},\,{}^{3}D_{2}$, carrying the first
$\Mthree$. The $^{3}P_{2}$ electric dipole is the first correction to the single
$\Eone$ contact of Eq.~\eqref{eq:cE1P0}. Across the nucleosynthesis and photodisintegration ranges, all of these effects are safely negligible.

\section{Cross Section}
\label{sec:xsec}

\subsection{Multipole Structure}

The tree amplitude to leading boundary order is
$\mathcal{M}=\mathcal{M}^{\mathrm{fact}}+\mathcal{B}_W
+\frac{e}{\Lambda^{3}}\sum_a c_a\mathcal{B}_a$. We square with the initial spin average $\tfrac14$ for the
incoming nucleons, sum over deuteron polarizations, and multiply by the flux and phase space factors.
Integrating over the scattering angle results in an incoherent sum over multipoles.

The cross section takes the form
\begin{equation}
  \sigma(np\to d\gamma)=\frac{\alpha(\gamma^2+\prel^2)}{4\prel}
  \bigg[\,2\,\big|\mathcal{A}_{\Mone}\big|^{2}
   +\sum_{J=0}^{2}\big|\mathcal{A}_{\Eone}^{(J)}\big|^{2}\,\bigg],
  \label{eq:sigma}
\end{equation}
with $\prel=\sqrt{m_N E_{\mathrm{cm}}}$ the relative momentum of the incoming nucleons,
$\gamma=\sqrt{m_N B}=45.7\,\mathrm{MeV}$ the deuteron binding momentum and
$B=2.224\,\mathrm{MeV}$ the binding energy. Because the initial state feels no Coulomb repulsion, the astrophysical $S$-factor is simply related to the cross section by $S(E) = E\,\sigma(E)$.

The two multipoles enter incoherently, with spin sums $2$ and
$\sum_{J}w_{J}^{2}=8$, and tree amplitudes:
\begin{align}
  \mathcal{A}_{\Mone} &= \sqrt{2m_{d}}\;\Egam\,U_{\Mone},
  \label{eq:AM1tree}\\
  \mathcal{A}_{\Eone}^{(J)} &= \sqrt{2m_{d}}\;
     \frac{w_{J}\,\prel}{\gamma^{2}+\prel^{2}}\,U_{\Eone},
  \label{eq:AE1tree}
\end{align}
with $w_{J}=\sqrt{8(2J+1)/9}$ for ${}^{3}P_{0,1,2}$. Their reduced matrix
elements are:
\begin{align}
  U_{\Mone} &= \frac{\CSv}{m_{d}}\,(1+\kappaV)\,P_N
               + \frac{c_{\Mone}}{\Lambda^{3}}, \label{eq:UM1np}\\
  U_{\Eone} &= \frac{e_{\mathrm{eff}}}{e}\,\frac{\CSv}{m_{d}}
               + \frac{\Egam^{2}\,c_{\Eone}}{\Lambda^{3}}, \label{eq:UE1np}
\end{align}
where $\CSv$ is the deuteron vertex of Eq.~\eqref{eq:strongSD} taken on shell,
i.e.\ at the deuteron pole. The residue matching of Eq.~\eqref{eq:residue} fixes
it to $\CSv^{\mathrm{ANC}}=\AS\sqrt{2\pi m_{d}}/m_{N}$, which
Appendix~\ref{app:bubble} recovers from the deuteron's wavefunction
renormalization. Here $P_N=1/(2m_N\Egam)$ is the threshold nucleon propagator, the contact terms
are isovector components with order-one constants of the multipole projection absorbed,
\begin{equation}
  e_{\mathrm{eff}} = e\,\frac{Z_{n}m_{p}-Z_{p}m_{n}}{m_{n}+m_{p}}
                   = -\,e\,\frac{m_{n}}{m_{n}+m_{p}} \simeq -0.500\,e ,
  \label{eq:eeff}
\end{equation}
is the recoil charge of the relative $n$--$p$ coordinate, and
\begin{equation}
  (1+\kappaV) \;=\; \mu_p-\mu_n \;\simeq\; 4.706 ,
  \label{eq:isovector}
\end{equation}
is the isovector part of the magnetic operator.

\subsection{Initial-State Rescattering}
\label{sec:isi}

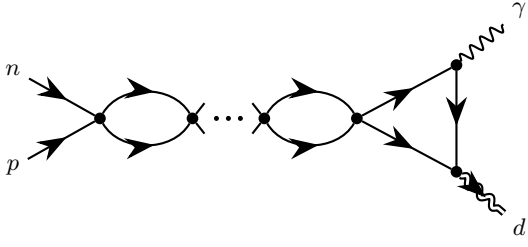
\begin{figure}[t]
\centering
\begin{tikzpicture}[fmg, every node/.append style={font=\small}, scale=0.72]
  \path (-1.6,0.9) node (n) {$n$} (-1.6,-0.9) node (p) {$p$}
        (0,0) node[vertex,name=C1] {};
  \draw[nucl] (n) -- (C1); \draw[nucl] (p) -- (C1);
  \path (1.7,0) node[vertex,name=C2] {};
  \draw[nucl] (C1) to[bend left=60] (C2);
  \draw[nucl] (C1) to[bend right=60] (C2);
  \draw[thick] (C2) -- (1.92,0.28); \draw[thick] (C2) -- (1.92,-0.28);
  \foreach \x in {2.14,2.36,2.58} \fill (\x,0) circle (1.3pt);
  \path (3.02,0) node[vertex,name=C3] {};
  \draw[thick] (2.80,0.28) -- (C3); \draw[thick] (2.80,-0.28) -- (C3);
  \path (4.72,0) node[vertex,name=C4] {};
  \draw[nucl] (C3) to[bend left=60] (C4);
  \draw[nucl] (C3) to[bend right=60] (C4);
  \path (6.55, 0.98) node[vertex,name=Vg] {};
  \path (6.55,-0.98) node[vertex,name=Vd] {};
  \draw[nucl] (C4) -- (Vg);
  \draw[nucl] (Vg) -- (Vd);
  \draw[nucl] (C4) -- (Vd);
  \path (7.70, 1.98) node (g) {$\gamma$};
  \draw[phot] (Vg) -- (g);
  \path (7.70,-1.98) node (d) {$d$};
  \dwave{Vd}{d}
\end{tikzpicture}
\caption{Schematic diagram showing rescattering of the initial state prior to the fusion process.}
\label{fig:bubblechain}
\end{figure}

Both multipoles reach the deuteron through a strongly interacting $np$ pair at low relative momentum:
the two nucleons rescatter through short-range interactions any number of times before fusing.
In the EFT this is the ladder of bubble diagrams of Fig.~\ref{fig:bubblechain}.
Resumming all of them converts the plane wave into the full $^{1}S_{0}$ scattering state, the Watson
enhancement~\cite{Watson:1954uc} of the entrance channel.
The dominant near-threshold transition is the isovector $\Mone$ from an incoming $^{1}S_{0}$ wave
to the $^{3}S_{1}$ deuteron. There the enhancement is large: a fine-tuned scattering length places a virtual state only
$66\,\mathrm{keV}$ below threshold. Each rescattering is an $\mathcal{O}(1)$
effect and the ladder must be summed to all orders, producing a $^{1}S_{0}$ effective-range denominator that dresses the magnetic
amplitude.
The $\Eone$ multipole is
different, proceeding from the $^{3}P_{J}$ incoming waves, with no nearby pole and no
anomalously large scattering volume. Its ladder is resummed in the same way, but what
comes out is a small correction rather than an $\mathcal{O}(1)$ enhancement.

Rescattering is treated in detail in Appendix~\ref{app:bubble}. It dresses the
tree amplitudes of Eqs.~\eqref{eq:AM1tree}--\eqref{eq:AE1tree}, which take the modified forms:
\begin{align}
  \mathcal{A}_{\Mone} &= \sqrt{\frac{2}{m_{d}}}\;\frac{1}{8\pi} T_{^1S_0}
    \Big[(1+\kappaV)\,\CSv\,W(\prel^{2}) \nonumber\\
   &\qquad\qquad\qquad +(\gamma^{2}+\prel^{2})\,\ell_{1}\Big],
  \label{eq:AM1}\\
  \mathcal{A}_{\Eone}^{(J)} &= \sqrt{\frac{2}{m_{d}}}\;
    \frac{w_{J}\,\prel\,m_{d}}{\gamma^{2}+\prel^{2}}
    \bigg[\frac{e_{\mathrm{eff}}}{e}\,\frac{\CSv}{m_{d}}\,
      \big(1+\delta_{\Eone}^{(J)}\big) \nonumber\\
   &\qquad\qquad\quad
    +\,\delta_{J0}\;\frac{9\,\Egam^{2}\,c_{\Eone}}{\Lambda^{3}}\,
      e^{i\delta_{^{3}P_{0}}}\cos\delta_{^{3}P_{0}}\bigg],
  \label{eq:AE1}
\end{align}
the magnetic amplitude reducing to its tree form when the nucleon four-point
interaction is switched off, and the electric one likewise up to the contact's
localization in ${}^{3}P_{0}$---the factor of nine and the single surviving channel
are derived below [Eq.~\eqref{eq:cE1P0}], and the contact term's dressing scheme is
fixed in Appendix~\ref{app:E1dress}. Here $T_{^1S_0}$ is the elastic rescattering amplitude,
Eq.~\eqref{eq:NRform},
\begin{equation}
  T_{^1S_0} = \frac{4\pi/m_{N}}
    {-1/a_{0}+\tfrac12 r_{0}\prel^{2}-i\prel}\,,
  \label{eq:T1S0}
\end{equation}
in terms of the ${}^{1}S_{0}$ effective-range parameters $a_{0}$ and $r_{0}$, and
\begin{equation}
  W(\prel^{2}) = \gamma-\frac{1}{a_{0}}
    -\frac{(r_{0}+\rho_{d})\,\gamma^{2}}{4}
    +\frac{(r_{0}-\rho_{d})\,\prel^{2}}{4}
  \label{eq:Wdef}
\end{equation}
is the impulse overlap.

As discussed in Appendix~\ref{app:bubble}, the deuteron parameter $\rho_{d}$ appearing in $W$ is not an independent parameter, but just (given $\gamma$)
a rewriting of the ANC,
\begin{equation}
  1-\gamma\rho_{d} = \frac{2\gamma}{\AS^{2}}\,,
  \label{eq:rhoANC}
\end{equation}
representing the deuteron on-shell residue in effective-range
units. It appears only because of the scheme choice relating the two-body
current $\ell_{1}$ of Eq.~\eqref{eq:AM1} to the renormalized coefficient
$c_{\Mone}$,
\begin{equation}
  \ell_{1} \equiv \frac{2m_{d}W(0)}{\Lambda^{3}}\,c_{\Mone},
  \label{eq:cM1def}
\end{equation}
with $W(0)$ the threshold value of Eq.~\eqref{eq:Wdef}. The scheme is convenient
because $\ell_{1}$ is the form used in the dibaryon
literature~\cite{Ando:2005cz,Rupak:1999rk} while $c_{\Mone}$ is the coefficient
informed by naive dimensional analysis; nothing in the amplitude depends on the choice.

The electric sector carries a single constant of its own, $c_{\Eone}$, entering
Eq.~\eqref{eq:UE1np} with the soft-photon factor $\Egam^{2}$. Both contact terms are
dimensionless and $\mathcal{O}(1)$ by naive dimensional
analysis~\cite{Manohar:1983md}, so their
fitted values may be compared directly against that expectation.

The factor $\delta_{\Eone}^{(J)}$ in Eq.~\eqref{eq:AE1} is the $^{3}P_{J}$
initial-state interaction, the one correction to the electric amplitude that the
EFT resolves at this order. Each entrance channel is dressed by a single insertion
of its own elastic amplitude, which for a contact interaction is the whole ladder,
so that
\begin{equation}
  \delta_{\Eone}^{(J)} = \sin\delta_{^{3}P_{J}}\,e^{i\delta_{^{3}P_{J}}}
      \left(\frac{\gamma\big(\gamma^{2}+3\prel^{2}\big)}{2\prel^{3}}+i\right),
  \label{eq:dE1}
\end{equation}
derived in Appendix~\ref{app:E1dress}, with $\delta_{^{3}P_{J}}$ the measured
Nijmegen phases~\cite{Stoks:1993tb,PavonValderrama:2004se}. The coefficient of $i$
is fixed to unity by unitarity, and with it the dressed amplitude
\begin{equation}
  1+\delta_{\Eone}^{(J)} = e^{i\delta_{^{3}P_{J}}}
    \left(\cos\delta_{^{3}P_{J}}
      +\frac{\gamma\big(\gamma^{2}+3\prel^{2}\big)}{2\prel^{3}}\,
       \sin\delta_{^{3}P_{J}}\right)
  \label{eq:dE1watson}
\end{equation}
carries the elastic ${}^{3}P_{J}$ phase and nothing more, as Watson's theorem
requires. Both terms in Eq.~\eqref{eq:dE1} are half the coefficient printed in
Refs.~\cite{Chen:1999bg,Rupak:1999rk}, whose amplitude-level correction
$(m_{N}\gamma/12\pi)(\gamma^{2}/3+\prel^{2})\,C_{p}$ gives a production amplitude
of phase $2\delta_{^{3}P_{J}}$. The spurious half of that term is at most $0.2\%$
of $\sigma_{\Eone}$ over the capture range where those works quote their accuracy,
and $0.6\%$ at the top of the photodisintegration window fitted in
Ref.~\cite{Rupak:1999rk}.

Expanded at small phase, Eq.~\eqref{eq:dE1} is
$\delta_{\Eone}^{(J)}=-\tfrac{3}{2}\gamma\big(\gamma^{2}/3+\prel^{2}\big)a_{1}^{(J)}$ in terms of
the scattering volumes, and its $(2J+1)/9$ average is the single constant $C_{p}$ of
Eq.~\eqref{eq:Cpsum}. We do not use that average. It is a twelvefold cancellation
between the attractive $^{3}P_{0,2}$ and the repulsive $^{3}P_{1}$, and the three
channels carry very different effective ranges, so a cancellation fixed at threshold
does not survive the fitted range. At $\Egam=7.1\,\mathrm{MeV}$ the channel
corrections are $+7.0\%$, $-4.1\%$ and $+1.4\%$ while the factor they produce in
$\sigma_{\Eone}$ is $+0.46\%$, and the effective $C_{p}$ changes sign near
$\Egam\simeq4\,\mathrm{MeV}$. 

Two further sub-percent effects are deliberately omitted. The deuteron $D$ state
contributes $-5\eta_{d}^{2}(1-\gamma\rho_{d})$, a constant $-0.19\%$ of $\sigma_{\Eone}$; and the
relativistic corrections to the amplitude and to the two-body phase space
together supply between $-0.27\%$ and $-0.65\%$ of it across the same range. Both are flat enough in energy to be nearly degenerate with the contact
$c_{\Eone}$ that multiplies the same amplitude. Dropping them shifts
$\sigma(E)$ by less than $0.1\%$ everywhere between threshold and
$1\,\mathrm{MeV}$, and the reaction rate by less than $0.1\%$ across
$T_{9}=0.01$--$3$. That is the largest of the omitted classes, and it sits a factor
of two inside the band quoted in Sec.~\ref{sec:results}; it is collected there with
the rest of the truncation.

The multipoles absent from Eq.~\eqref{eq:sigma} are smaller still. The leading
one-body electric quadrupole proceeds from the $d$-wave entrance, with recoil
charge $e/4$ and two further long-wavelength powers of $\Egam$; in the zero-range
limit it contributes $1\times10^{-3}$ of $\sigma_{\Eone}$ at the top of the SLEGS
range and at most $10^{-4}$ of the cross section across the nucleosynthesis
window. The magnetic quadrupole, which trades the recoil charge for the nucleon
moments at a further order of the recoil expansion, is smaller by another factor
of several. Neither approaches the truncation budget of Sec.~\ref{sec:results}.

The $\Delta_{W}=7$ contact terms reach only one entrance channel. That sector is rank two,
with $2c_{1}-c_{2}-2c_{3}$ the $^{3}P_{0}$ direction and $c_{4}-c_{5}$ the
$^{3}P_{1}$ direction, and a Jacob--Wick projection shows the second is pure $M2$ up to a recoil correction of relative size $\Egam/m_{N}$; the $^{3}P_{2}$
channel is empty altogether, and opens only at the next order. So
\begin{equation}
  c_{\Eone}^{(1)}=c_{\Eone}^{(2)}=0,
  \qquad c_{\Eone}=\sum_{J}\frac{2J+1}{9}\,c_{\Eone}^{(J)}=\frac{1}{9}c_{\Eone}^{(0)},
  \label{eq:cE1P0}
\end{equation}
and the single electric constant we fit is the $^{3}P_{0}$ constant, nothing else.
$c_{\Eone}^{(0)}$ is renormalized by divergences in the rescattering ladder, with the scheme fixed in Appendix~\ref{app:E1dress} 
such that only the Watson phase $e^{i\delta_{^{3}P_{0}}}\cos\delta_{^{3}P_{0}}$ is kept (see Eq.~\eqref{eq:AE1}). 
The dimensional-analysis expectation quoted above is calibrated on the $J$-averaged
amplitude, so $c_{\Eone}$ rather than $c_{\Eone}^{(0)}$ is the relevant object. 

\section{Results}
\label{sec:results}

We confront the cross section of Sec.~\ref{sec:xsec} with data, aiming to produce the best-informed
description of $\sigma(E)$ and $S(E)$ consistent with theoretical inputs and experimental measurements.
Everything the cross section depends on at the percent level or higher is included as a fit parameter in a single Bayesian analysis of the thermal
capture cross section together with the deuteron photodisintegration dataset of
SLEGS. The cross section, the thermonuclear reaction rate, and the primordial
deuterium shift are read from the resulting posterior.

\subsection{Joint Fit}
\label{sec:jointfit}

The fit parameters are:
\begin{equation}
  \theta=\big\{\,\CSv,\;a_{0},\;r_{0},\;c_{\Mone},\;c_{\Eone},\;\lambda_{S}\,\big\},
  \label{eq:theta}
\end{equation}
including the deuteron vertex coupling, the two ${}^{1}S_{0}$ effective-range parameters
of Eq.~\eqref{eq:T1S0}, two contact coefficients, and the overall normalization
$\lambda_{S}$ of the photodisintegration data.

We sample the log-posterior
\begin{equation}
\begin{aligned}
  -2\ln\mathcal{P}(\theta)=\;
  & \sum_{k=1}^{2}\frac{\big[\sigma(E_{\mathrm{th}};\theta)-\sigth^{(k)}\big]^{2}}{\big(\delta\sigth^{(k)}\big)^{2}}\\
  & +\sum_{i=1}^{22}
    \frac{\big[\,\sigma_{i}^{\mathrm{fold}}(\theta)-\sigma_{i}^{\mathrm{f}}\,\big]^{2}}
         {\big(\delta\sigma_{i}^{\mathrm{f}}\big)^{2}}\\
  & +\left(\frac{\lambda_{S}-1}{\delta_{S}}\right)^{2}
    +\left(\frac{\CSv-\CSv^{\mathrm{ANC}}}{\sigma_{\CSv}}\right)^{2} \\
  & +\left(\frac{a_{0}-\bar a_{0}}{\sigma_{a_{0}}}\right)^{2}
    +\left(\frac{r_{0}-\bar r_{0}}{\sigma_{r_{0}}}\right)^{2},
\end{aligned}
  \label{eq:logpost}
\end{equation}
where $f$ is the detailed-balance factor necessary to convert the fusion process into the photodisintegration cross section,
\begin{equation}
f \equiv \frac{\sigma(\gamma d\to np)}{\sigma(np\to d\gamma)}
  =\frac{2}{3}\,\frac{\prel^{2}}{k^{2}},
  \label{eq:detbal}
\end{equation}
the $2/3$ being the ratio of initial spin weights.  We use exact two-body
kinematics throughout: for a photon of laboratory energy $\Egam$ incident on a
deuteron at rest,
\begin{equation}
  s=m_{d}^{2}+2m_{d}\Egam,\quad
  \prel^{2}=\frac{\lambda\big(s,m_{n}^{2},m_{p}^{2}\big)}{4s},\quad
  k^{2}=\frac{\big(s-m_{d}^{2}\big)^{2}}{4s},
  \label{eq:exactkin}
\end{equation}
with $\lambda$ the Källén function, and the capture cross section is evaluated at
the corresponding exact center-of-mass energy $E_{\mathrm{cm}}=\prel^{2}/m_{N}$.
The recoil expansion of Eq.~\eqref{eq:detbal} is accurate in $f$ itself to better
than $0.01\%$, but it displaces $E_{\mathrm{cm}}$ by $1.4\%$ at the lowest beam
energy, which moves that folded point by $1.9\%$; it also turns negative $1.3\,$keV
above threshold, where the exact form simply vanishes at the physical threshold
$\Egam=[(m_{n}+m_{p})^{2}-m_{d}^{2}]/2m_{d}$.
The photodisintegration term compares against the \emph{folded} cross sections
$\sigma_{i}^{\mathrm{f}}$ the experiment measures. The theory is therefore
convolved with the normalized beam spectrum $D_{i}$ of the $i$th collision angle, over the
same interval the experiment integrates,
\begin{equation}
  \sigma_{i}^{\mathrm{fold}}(\theta)=\lambda_{S}\!
    \int_{S_{n}}^{E_{\mathrm{max}}^{(i)}}\!\!
    D_{i}(\Egam)\,f(\Egam)\,\sigma\big(E_{\mathrm{cm}}(\Egam);\theta\big)\,d\Egam ,
  \label{eq:fold}
\end{equation}
so the detailed-balance factor sits inside the integral rather than being evaluated once
per point. The upper limit $E_{\mathrm{max}}^{(i)}$ is the kinematic edge of the
Compton-scattered beam. The normalization $\lambda_{S}$ is common to all the
photodisintegration data and carries the collaboration's quoted systematic
$\delta_{S}=3.6\%$ as its width. The first two lines are the likelihood; the last two
are the priors, on the normalization and on the three strong-interaction inputs. 
These entries are specified in detail in Sec.~\ref{sec:inputs}, and collected for
convenience in Table~\ref{tab:fitinputs}.

Eq.~(\ref{eq:logpost}) treats the strong-interaction inputs as parameters, free to move within their own
uncertainties as the production and photodisintegration data demand, and this
freedom propagates into the derived quantities through the same posterior. This
matters because the posterior correlates them: for example, the thermal
cross section fixes the magnetic amplitude at threshold, and $r_{0}$ decides how
much of that amplitude belongs to the impulse contribution and how much is generated by the boundary terms. 
Neither the pionless benchmark of Ref.~\cite{Rupak:1999rk} nor the dibaryon calculation of
Ref.~\cite{Ando:2005cz} are included in the fit; both are treated as independent
comparisons in Sec.~\ref{sec:consequences}.

The posterior $\propto e^{-\chi^{2}/2}$ is sampled with an affine-invariant
ensemble sampler~\cite{Goodman:2010dyf,Foreman-Mackey:2012any}: $48$ walkers
advanced $800$ steps and then $5000$, thinned by ten, at a mean acceptance
fraction of $0.52$. Parameters are quoted as posterior medians with $68\%$
intervals.

\subsection{Inputs}
\label{sec:inputs}

\begin{table}[t]
  \centering
  {\scriptsize
  \begin{tabular}{lcc}
    \hline\hline
    Quantity & Source & Value \\
    \hline
    \multicolumn{3}{l}{\bf Data} \\
    $\sigth$ ($2200$ m/s)    & Thermal capture~\cite{Cox1965} & $334.2\pm0.5$ mb \\
    $\sigth$ ($2200$ m/s)    & Thermal capture~\cite{Cokinos:1977zz} & $332.6\pm0.7$ mb \\
    $\sigma^{\mathrm{f}}(\gamma d{\to}np)$ & SLEGS $22$ points~\cite{Chen:2025oyj} & $2.33$--$7.09$ MeV \\
    $\lambda_{S}$            & SLEGS normalization~\cite{Chen:2025oyj} & $1.000\pm0.036$ \\
    \hline
    \multicolumn{3}{l}{\bf Theory} \\
    $\CSv$                   & ANC $\AS$~\cite{deSwart:1995ui} & $1.4367\pm0.0015$ \\
    $a_{0}$ [fm]             & $np$ ${}^{1}S_{0}$~\cite{Stoks:1993tb} & $-23.715\pm0.020$ \\
    $r_{0}$ [fm]             & PWA93~\cite{Stoks:1993tb} & $+2.706\pm0.060$ \\
    \hline
    \multicolumn{3}{l}{\bf Fixed} \\
    $1+\kappaV$              & ~\cite{ParticleDataGroup:2024cfk} & $4.706$ \\
    $e_{\mathrm{eff}}/e$     & ~\cite{ParticleDataGroup:2024cfk} & $-0.500$ \\
    $\Lambda$                & Eq.~\eqref{eq:Lambda} & $162.0$ MeV \\
    $a_{1}^{(J)}$ [fm$^{3}$] & NijmII~\cite{PavonValderrama:2004se} & $-2.468,\,1.529,\,-0.2844$ \\
    \hline
    \multicolumn{3}{l}{\bf Fit} \\
    $c_{\Mone}$              & Flat prior        & Isovector $\Mone$ \\
    $c_{\Eone}$              & Flat prior        & $\Eone$  \\
    \hline\hline
  \end{tabular}}
  \caption{Inputs to the joint fit of Eq.~\eqref{eq:logpost}, including: experimental measurements; fixed inputs with negligible uncertainties;
 and theory parameters which are fit subject to the indicated Gaussian uncertainties. The
 photodisintegration entry is the folded cross section measured at each of the twenty-two
 beam settings; the overall normalization $\lambda_{S}$ is fit subject to the
 collaboration's quoted common systematic.}
  \label{tab:fitinputs}
\end{table}

\subsubsection{Data}
\label{sec:data}

Two experimental measurements drive the fit, constraining different EFT
parameters. The thermal capture cross section at neutron velocity
$2200\,\mathrm{m/s}$ sits at an energy where the magnetic dipole term is
essentially the entire cross section; it anchors the fit to $c_{\Mone}$. It has
been measured twice, $\sigth=334.2\pm0.5\,\mathrm{mb}$~\cite{Cox1965} and
$332.6\pm0.7\,\mathrm{mb}$~\cite{Cokinos:1977zz}, and we carry both as separate
terms in Eq.~\eqref{eq:logpost}. The two differ by $1.9\sigma$, giving
$\chi^{2}=3.5$ for one degree of freedom, so we scale both uncertainties by the
factor $S=\sqrt{\chi^{2}/\mathrm{dof}}=1.86$ prescribed by the Particle Data
Group for discrepant input data~\cite{ParticleDataGroup:2024cfk}. The pair thus
constrains the thermal cross section to $333.7\pm0.8\,\mathrm{mb}$; without the
scaling it would constrain it to $\pm0.4\,\mathrm{mb}$, tighter than either
measurement alone.
This cross section has also been calculated: two chiral-EFT results that truncate the electromagnetic current before the
short-range isovector $\Mone$ operators sit $3.5$--$4\%$
low---$321.0\pm0.7$~\cite{Acharya:2021lrv} and
$322\pm3\,\mathrm{mb}$~\cite{Du:2022zds}---while the lattice-QCD calculation of
Ref.~\cite{Beane:2015yha}, which determines the corresponding two-body
counterterm from QCD rather than from data, gives
$332.4^{+5.4}_{-4.7}\,\mathrm{mb}$. 
These theoretical determinations are not included in the fit.

The twenty-two SLEGS photodisintegration cross sections~\cite{Chen:2025oyj}, spanning
$\Egam=2.327$--$7.089\,\mathrm{MeV}$ from
threshold through the peak and into the tail, are related to the capture
channel by crossing through detailed balance [Eq.~\eqref{eq:detbal}]; their energy
dependence is largely what constrains $c_{\Eone}$. Each enters with its statistical and
methodological errors added in quadrature, the $3.6\%$ scale systematic being carried
by $\lambda_{S}$ instead. Each is a
\emph{folded} observable, measured at one beam setting: the quasi-monochromatic
$\gamma$ beam carries a finite width, and the experiment records the cross section
convolved with that beam's energy distribution. We fit the folded cross sections directly,
convolving the theory with the corresponding beam spectrum and detailed-balance factor. The overall normalization floats
with the quoted $3.6\%$ common systematic as its prior, returning
$\lambda_{S}=0.981\pm0.006$.

\subsubsection{Strong-Interaction Inputs}

Three quantities are well determined externally, with uncertainties that move
the amplitude at a level comparable to those of the data. The deuteron vertex coupling
is tied to the asymptotic normalization through Eq.~\eqref{eq:residue},
$\AS=0.8846(9)\,\mathrm{fm}^{-1/2}$~\cite{deSwart:1995ui}, giving
$\CSv=1.4367\pm0.0015$; the range parameter $\rho_{d}$ follows from it by
Eq.~\eqref{eq:rhoANC} and is not independently varied. The ${}^{1}S_{0}$
scattering length and effective range are included as measurements
$a_{0}=-23.715\pm0.020\,\mathrm{fm}$ and $r_{0}=+2.706\pm0.060\,\mathrm{fm}$,
the width of the latter set by the spread across partial-wave analyses rather
than by any single determination. All three remain at their central values in the
fit, but their uncertainties propagate into the contact coefficients.

The two boundary coefficients, $c_{\Mone}$ and $c_{\Eone}$, are left free to adjust themselves with flat
priors [-30.0, +30.0].  Their fit values provide a measure of how well naive dimensional analysis power counting succeeds
at describing the EFT higher order terms.

\subsubsection{Fixed Inputs}

The masses, the binding energy, the isovector moment $1+\kappaV=4.706$, and the
recoil charge $e_{\mathrm{eff}}=-0.500\,e$ follow from quantities measured to far
better than our working precision. The ${}^{3}P_{J}$ phases enter only through
$\delta_{\Eone}^{(J)}$ [Eq.~\eqref{eq:dE1}], whose effect on $\sigma_{\Eone}$ stays
at or below the permille level across the nucleosynthesis window and reaches
$0.46\%$ at the highest beam. They are held fixed at their measured values, taken from the
effective-range form below $T_{\mathrm{lab}}=5\,\mathrm{MeV}$, where the tabulated
phases are quoted to too few digits to recover the volumes, and from the tabulation
above it; the sensitivity of the fit to that choice is quoted with $c_{\Eone}$ in
Sec.~\ref{sec:results}.

\subsection{Fit and EFT Parameters}
\label{sec:params}

\begin{table}[t]
  \centering
  \begin{tabular}{lcc}
    \hline\hline
    Parameter & Prior & Posterior \\
    \hline
    $\CSv$ & $1.4367\pm0.0015$~\cite{deSwart:1995ui} & $1.4369\pm0.0015$ \\
    $a_{0}$ [fm] & $-23.715\pm0.020$ & $-23.717\pm0.020$ \\
    $r_{0}$ [fm] & $+2.706\pm0.060$ & $+2.712\pm0.060$ \\
    $c_{\Mone}$ & Flat & $+0.162\pm0.015$ \\
    $c_{\Eone}$ & Flat & $-0.88\pm0.21\,{}^{+0.16}_{-0.08}\pm0.03$ \\
    $\lambda_{S}$ & $1.000\pm0.036$~\cite{Chen:2025oyj} & $0.981\pm0.006$ \\
    \hline
    $\rho_{d}$ [fm] & Derived & $+1.763\pm0.005$ \\
    $\ell_{1}$ [fm] & Derived & $+1.19\pm0.11$ \\
    $L_{\Eone}$ [fm$^{3}$] & Derived & $+14.9\pm3.5\,{}^{+1.3}_{-2.6}\pm0.5$ \\
    $g_{A}$ [$10^{4}$] & Derived & $+3.074\pm0.067$ \\
    $g_{B}$ [$10^{4}$] & Derived & $-3.070\pm0.067$ \\
    $g_{A}+g_{B}$ & Derived & $+42.22\pm0.04$ \\

    \hline\hline
  \end{tabular}
  \caption{Joint-fit posterior median values and $1\sigma$ uncertainties, including the fitted quantities (above the rule) and quantities derived from them (below). The second uncertainty on $c_{\Eone}$ and $L_{\Eone}$ is the systematic spanning the SLEGS treatments of Sec.~\ref{sec:params}, quoted as an asymmetric envelope about the primary fit. 
  The third spans the treatment of the ${}^{3}P_{J}$ initial-state interaction.}
  \label{tab:posterior}
\end{table}

\begin{figure}[!tb]
  \centering
  \includegraphics[width=\columnwidth]{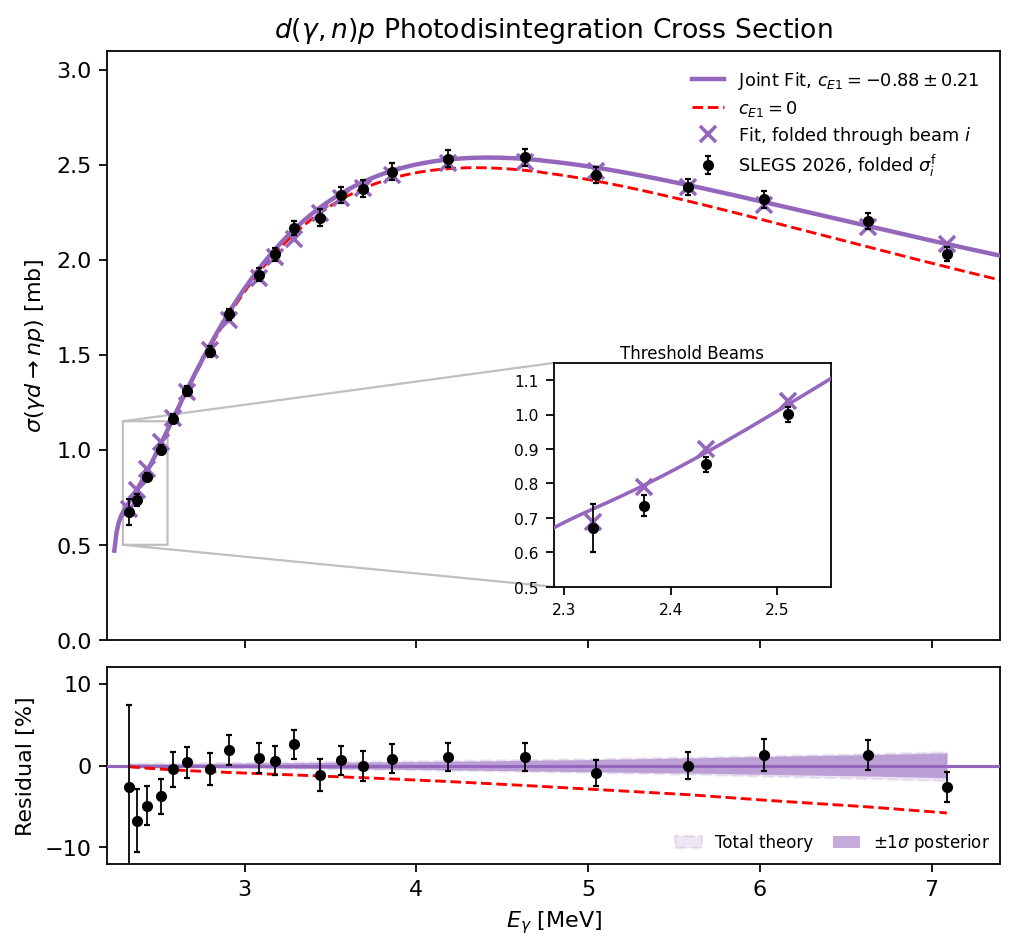}
  \caption{Crossed-channel deuteron photodisintegration $\sigma(\gamma d\to np)$.
  Circles are the SLEGS folded cross sections $\sigma^{\mathrm{f}}_{i}$, one per beam
  setting~\cite{Chen:2025oyj}; crosses are the fit folded through the same spectra;
  the solid curve is the monochromatic cross section, dashed with $c_{\Eone}=0$.
  Lower panel: residuals from the folded prediction, with the $\pm1\sigma$ posterior
  (solid) and the total theory uncertainty (dashed outline).}
  \label{fig:slegs}
\end{figure}

The posterior is collected in Table~\ref{tab:posterior}. At its maximum the data
$\chi^{2}$ is $19.9$ for the twenty-four points, with the three theory terms
contributing $0.04$ and the normalization prior on $\lambda_{S}$ a further $0.29$, for a
total of $20.3$. Each of the four priors is counted as a datum and each of the six
parameters as a degree of freedom, leaving $22$. The fit is comfortable, and it does not strain any of the
inputs.  Also shown are the derived quantities, including the dimensionless ${}^{1}S_{0}$ four-nucleon couplings
 $g_{A}$ and $g_{B}$ defined in Eq.~\eqref{eq:Csinglet}, obtained from $(a_{0},r_{0})$ by inverting
  Eq.~\eqref{eq:EREsinglet}. They are almost exactly anticorrelated, each of order $3\times10^{4}$ while the threshold combination
  they control is $\sim 42$.  $\ell_{1}$ and $L_{\Eone}$ are the two contacts in the dibaryon normalizations of Refs.~\cite{Ando:2005cz,Rupak:1999rk}, related to the fit coefficients by
  Eq.~\eqref{eq:cM1def} and $L_{\Eone}=-16.98\,c_{\Eone}\,\mathrm{fm}^{3}$.  Fig.~\ref{fig:slegs} shows the prediction for the crossed channel at the best-fit point and its $\pm 1 \sigma$
  error posterior band compared to the SLEGS data, folded through each beam spectrum (as the fit does).
  
\subsubsection{$c_{\Mone}$}

The thermal cross section fixes $c_{\Mone}=+0.162\pm0.015$. Setting $c_{\Mone}=0$ with every other parameter held at the best-fit point
would result in $306\pm2\,\mathrm{mb}$, or
$91.7\pm0.8\%$ of the fitted $333.5\pm0.8\,\mathrm{mb}$. This is the familiar statement that single-particle currents
under-predict thermal capture by about ten percent, historically attributed to
meson-exchange currents. Here one contact operator captures the physics,
\begin{equation}
  c_{\Mone} = +0.162\pm0.015 ,
  \label{eq:cM1fit}
\end{equation}
or $\ell_{1}=+1.19\pm0.11\,\mathrm{fm}$ in the dibaryon normalization of
Eq.~\eqref{eq:cM1def}. Its uncertainty follows from the range parameter, through the correlation
$\mathrm{corr}(r_{0},c_{\Mone})=+0.93$.
Naive dimensional analysis would suggest $c_{\Mone}=\mathcal{O}(1)$, and thus
this result is in harmony with EFT expectations. Strictly, the thermal cross
section fixes $c_{\Mone}$ together with whatever the next order contributes at
threshold, so the statement is one about the combination rather than about the
leading constant alone.

An exact threshold matching maps $c_{\Mone}$ onto the pionless-EFT
normalization, $L_{np}=-9.00\pm0.04\,\mathrm{fm}^{2}$, agreeing within
$0.8\sigma$ with the $-9.039\pm0.027\,\mathrm{fm}^{2}$ extracted by
Ref.~\cite{Rupak:1999rk} from the same data. The residual is a matter of
bookkeeping rather than physics. In the counting of Ref.~\cite{Rupak:1999rk} the
slope $\tilde{L}_{np}$ enters one order beyond $L_{np}$ and multiplies the same
function of $\prel$, so the thermal cross section fixes only the combination of
the two; our $c_{\Mone}$ encodes both.

The two chiral-EFT calculations inform the same
coefficient from the opposite direction. Both sit low---Ref.~\cite{Acharya:2021lrv}
by $3.8\%$ at $10\,\mathrm{keV}$, easing to $1.9\%$ at $1\,\mathrm{MeV}$, and
Ref.~\cite{Du:2022zds} by a flat $3.5$--$4.6\%$ in the $\Mone$ channel below
$10\,\mathrm{keV}$. The difference could be anticipated from the fact that both work with
electromagnetic currents truncated before the order at which the short-range
isovector $\Mone$ operators enter, and neither fits the thermal cross section:
they return $321$ and $322\,\mathrm{mb}$ at threshold against the measured
$333.7\pm0.8\,\mathrm{mb}$. Their quoted uncertainties, $0.2\%$ and $0.9\%$,
cover the truncation of the potential alone. What
their common offset measures is therefore the size of the two-body magnetic
current---the same physics the single constant $c_{\Mone}$ absorbs, and the same
ten percent the impulse approximation misses. It shrinks with energy because the
$\Eone$, which carries no such constant at leading order, takes over.

Comparing to Ref.~\cite{Du:2022zds} sharpens the picture, because they tabulate one-body and
two-body currents separately. Setting $c_{\Mone}=0$ in Eq.~\eqref{eq:AM1} extracts
the impulse amplitude, the object corresponding to their one-body contribution.  The two
show percent-level agreement: $306\pm2\,\mathrm{mb}$ compared to $311\pm3\,\mathrm{mb}$ at threshold,
and better than $2\%$ at every tabulated energy out to $10\,\mathrm{keV}$. 

\subsubsection{$c_{\Eone}$}

The fit returns a posterior for $c_{\Eone}$ of:
\begin{equation}
  c_{\Eone} = -0.88\pm0.21,
  \label{eq:cE1fit}
\end{equation}
essentially uncorrelated with everything else in the posterior. At this value, it raises $\sigma_{\Eone}$ by
$0.5\%$ at threshold and $2.5\%$ at the SLEGS peak. In the normalization of
Ref.~\cite{Rupak:1999rk} the same constant is
$L_{\Eone}=-16.98\,c_{\Eone}\,\mathrm{fm}^{3}=+14.9\pm3.5\,\mathrm{fm}^{3}$,
to be compared with the $-5.3\pm3.6\,\mathrm{fm}^{3}$ obtained there from the pre-2000
photodisintegration data. The SLEGS peak and tail sit above the $c_{\Eone}=0$ curve
where the older data sat below it.

The twenty-two folded cross sections are independent observables, one per beam setting,
and the fit uses them directly. The \emph{unfolded} monochromatic cross sections reported
alongside them are a different object: evaluations of a single eight-parameter function,
fitted so that its convolution with the beam spectra reproduces the folded data. Fitting
those instead would impose that parameterization on ours, so it is worth asking how much
the answer depends on the choice. Three treatments bracket it. Folding the beam spectra
supplied by the collaboration gives $-0.88\pm0.21$. Substituting our own reconstruction of
those spectra, traced from their Fig.~S1, gives $-0.95$. Fitting the unfolded values
instead, with the covariance obtained by propagating the folded uncertainties through
their eight-parameter form, gives $-0.72$. The three agree within the fit uncertainty, and
each returns a negative EFT coefficient of $\mathcal{O}(1)$--natural size. We attach the
spread as a treatment systematic, as the envelope
${}^{+0.16}_{-0.08}$ about the primary fit rather than as a symmetric band: the two
alternatives fall on the same side of it, and a symmetric error large enough to reach
one of them would overstate the spread in the other direction.

The treatment of the ${}^{3}P_{J}$ initial-state interaction supplies a second one.
Parameterizing the phases by the effective-range expansion rather than by the
Nijmegen tabulation shifts $c_{\Eone}$ by $0.03$, and adopting the scattering
volumes implied by the older potentials in place of the modern ones shifts it by
$0.03$. Those two are worth noting for what they are not: the threshold value of
$C_{p}$ is itself uncertain at the $20\%$ level, because it is the cancelling sum
of Eq.~\eqref{eq:Cpsum}, and yet moving it by that much moves $c_{\Eone}$ by only an
eighth of its fit uncertainty. Once the measured phases carry the energy
dependence, the threshold extrapolation stops controlling the answer. Propagated to
the observables the two together move $\sigma(E)$ by at most $0.03\%$ across the
nucleosynthesis window and the rate by $0.02\%$ at $T_{9}=1$ and $0.03\%$ at
$T_{9}=3$, and they are carried through to the rate band of
Sec.~\ref{sec:consequences} on that footing. Combining,
$c_{\Eone}=-0.88\pm0.21\,(\mathrm{fit})\,{}^{+0.16}_{-0.08}\,(\mathrm{treatment})
\pm0.03\,(^{3}P)$, or
$L_{\Eone}=+14.9\pm3.5\,{}^{+1.3}_{-2.6}\pm0.5\,\mathrm{fm}^{3}$.

The pull against Ref.~\cite{Rupak:1999rk} is thus $3.5\sigma$. The two determinations rest on disjoint data.
Ref.~\cite{Rupak:1999rk} fixed $L_{\Eone}$ from the photodisintegration measurements
available before 2000, which cluster near threshold, whereas the SLEGS set reaches
$7.1\,\mathrm{MeV}$ and constrains the $\Eone$ energy dependence directly. Since
$L_{\Eone}$ multiplies an $\Egam^{2}$ shape, that reach is precisely the lever arm the
older data lacked. It is not, however, carried by the highest-energy beams alone: cutting
the fit window at $5.6\,\mathrm{MeV}$ gives $c_{\Eone}=-1.11\pm0.38$, and every upper cut
we have tried leaves the pull between $3.0\sigma$ and $4.0\sigma$ while moving the
$T_{9}=1$ rate by at most $0.3\%$. The disagreement is therefore between the two data sets
rather than between the two calculations. It remains an
open issue for the $\Eone$ sector, and one the polarization observables of
Sec.~\ref{sec:discussion} bear on directly.

\subsubsection{$g_A$ and $g_B$}

The ${}^{1}S_{0}$ four-nucleon couplings corresponding to $(a_{0},r_{0})$ are nowhere near natural.  They are 
$g_{A}=+3.07\times10^{4}$ and $g_{B}=-3.07\times10^{4}$, reflecting the shallow
virtual state $66.3\,\mathrm{keV}$ below threshold. The threshold combination is
$g_{A}+g_{B}=4\pi\big(|a_{0}|\Lambda\big)\big(\Lambda/m_{N}\big)=42$, enhanced by
the scattering length, $|a_{0}|\Lambda=19.5$, whereas the individual couplings are enhanced by a further
$m_{N}^{2}|a_{0}|r_{0}/2=727$ from the $s_{np}$ slope of
Eq.~\eqref{eq:Csinglet} carrying an effective range of nuclear size.

At face value, couplings of this size would lead to a breakdown of the EFT
expansion at momenta of order $\Lambda/\sqrt{g_{A,B}}\sim1\,\mathrm{MeV}$, an
$E_{\mathrm{cm}}$ of about a keV---an order of magnitude below the bottom of the
nucleosynthesis window. 
Resumming their dominant contributions via the rescattering ladder mitigates this impact. 
What survives the resummation is the effective-range polynomial $\xi(\prel^{2})$ of Eq.~\eqref{eq:xidef}, whose coefficients are ordinary nuclear-scale numbers, and
the expansion should be judged by the importance of its successive terms rather than by the size of $g_{A,B}$ themselves. 
Its radius of convergence is set by the nearest singularity, the $t$-channel
pion cut at $\prel=m_{\pi}/2$, or $E_{\mathrm{cm}}=5.2\,\mathrm{MeV}$---the same
radius inside which Appendix~\ref{app:match} fits the elastic phases. 
The first omitted term, $v_{2}\simeq-1\,\mathrm{fm}^{3}$ (as estimated in Ref.~\cite{Rupak:1999rk}), 
moves the cross section by less than $0.03\%$ across the nucleosynthesis window and by $0.1\%$ at the top of the photodisintegration
range, where $\prel$ has reached $0.97\,(m_{\pi}/2)$ but the magnetic dipole has
fallen to $2.7\%$ of the total cross section. In other words, the correction to
the relevant multipole grows to a few percent at the edge of the fit region, but
the over-all importance of the multipole itself becomes tiny in that regime.

\subsection{Cross Section and Rate}
\label{sec:consequences}

\begin{figure}[t]
  \centering
  \includegraphics[width=\columnwidth]{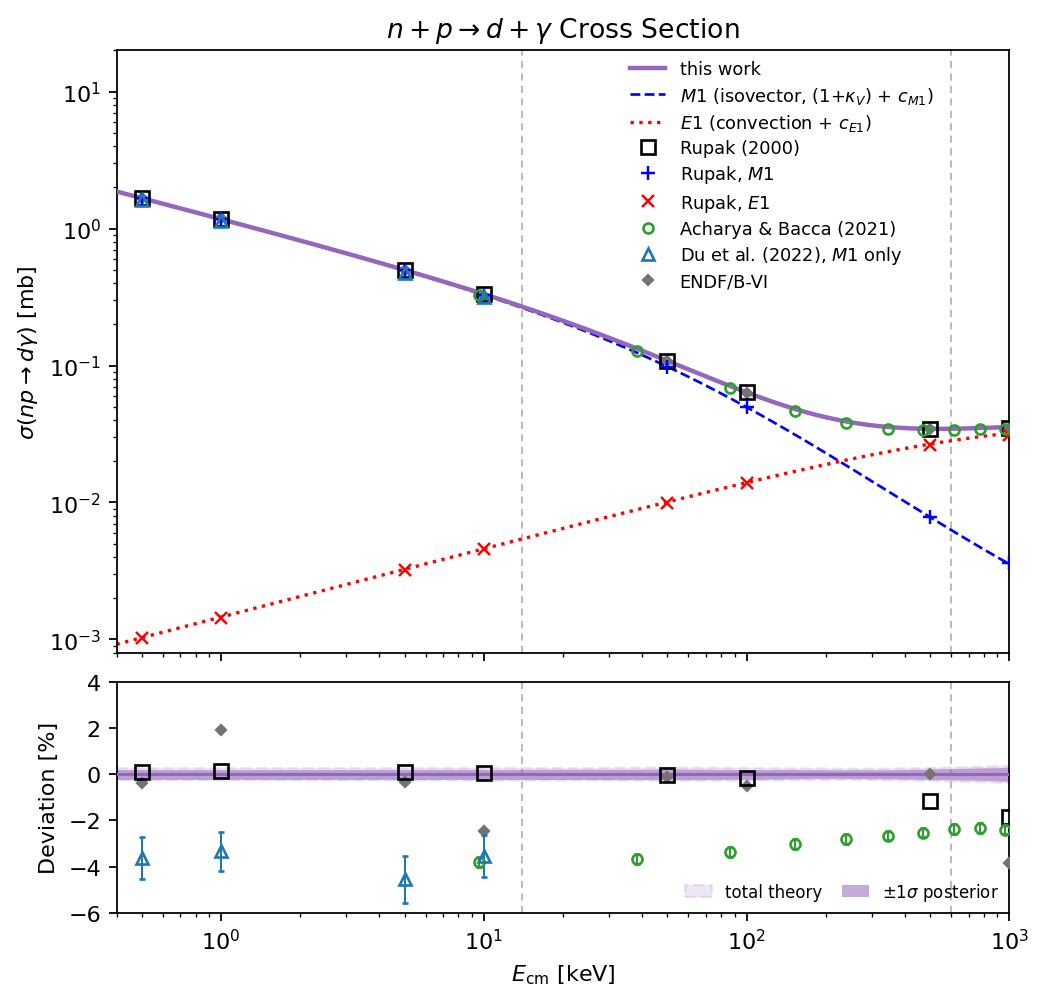}
  \caption{Cross section for $\npdg$ as computed in this work (solid purple), with
  the $\Mone$ (blue dashed) and $\Eone$ (red dotted) contributions at the
  best-fit parameters and the $\pm1\sigma$ posterior band (light purple). Also
  shown are the pionless-EFT determination of Ref.~\cite{Rupak:1999rk}
  (squares, crosses, and x's for the total, $\Mone$, and $\Eone$), chiral EFT
  from Refs.~\cite{Acharya:2021lrv} (circles) and~\cite{Du:2022zds} (triangles,
  $\Mone$ only), both with their quoted uncertainties, and the ENDF/B-VI
  compilation tabulated in Ref.~\cite{Rupak:1999rk} (diamonds). The dashed
  vertical lines mark the nucleosynthesis window, $14$--$600\,\mathrm{keV}$.
  Lower panel: deviation from our best fit, with the $\pm1\sigma$ posterior band
  (solid) and the total theory uncertainty including the treatment systematics and the
  truncation bound (dashed outline); the points from Ref.~\cite{Du:2022zds} show the
  deviation of their $\Mone$ from ours.}
  \label{fig:sigma}
\end{figure}

\begin{figure}[t]
  \centering
  \includegraphics[width=\columnwidth]{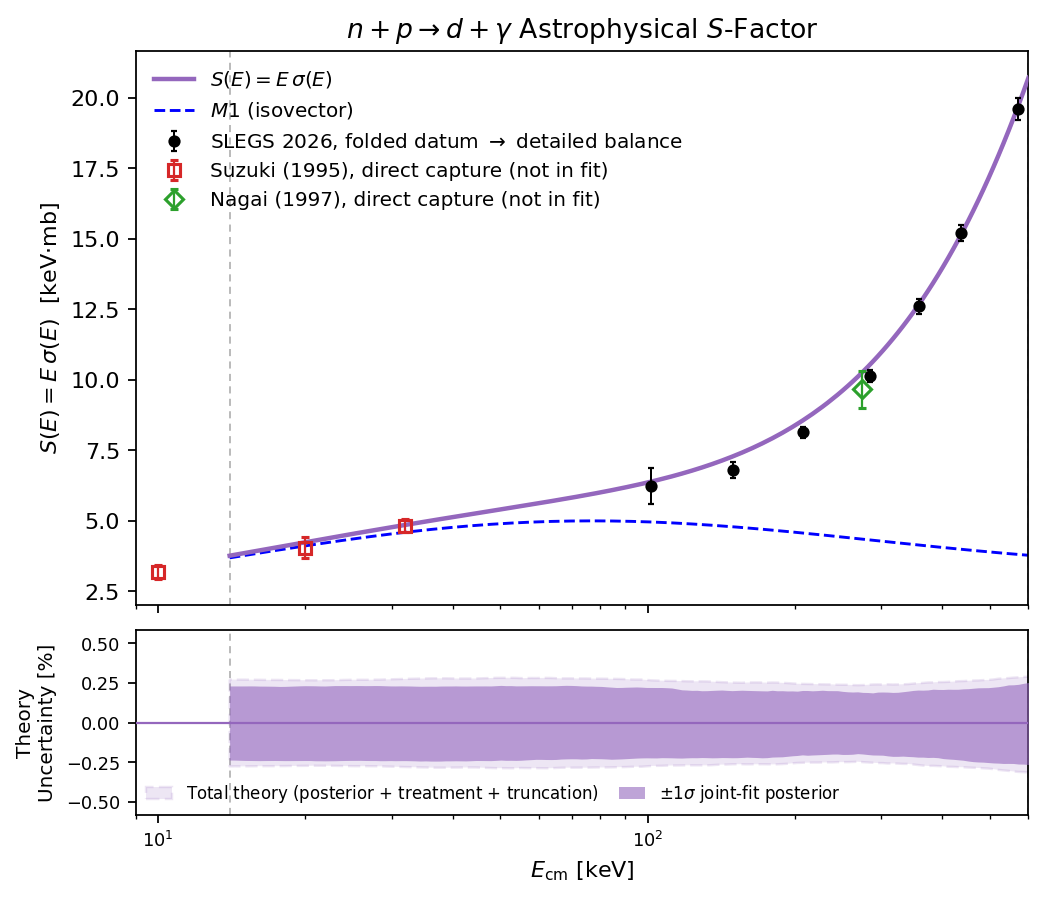}
  \caption{The $S$-factor $S(E)=E\,\sigma(E)$, showing the total (solid purple) and
  the isovector $\Mone$ (dashed blue). Also shown are the SLEGS 2026
  photodisintegration set~\cite{Chen:2025oyj} (filled circles), mapped onto this
  axis by the fit's own beam average, and the direct capture
  measurements of Refs.~\cite{Suzuki1995} (open squares) and~\cite{Nagai:1997zz}
  (open diamond). The dashed vertical line
  marks the lower edge of the nucleosynthesis window; its upper edge is the
  right-hand axis. The lower strip shows the theory uncertainty on an expanded
  scale: the $\pm1\sigma$ joint-fit posterior (solid) and the total including the
  treatment systematics and the truncation bound (dashed outline).}
  \label{fig:sfactor}
\end{figure}

Figure~\ref{fig:sigma} shows the cross section across the nucleosynthesis
window for all parameters at their best-fit values, together with the $\pm1\sigma$ posterior band. 
The isovector $\Mone$ falls as $1/v$ and dominates below
$E_{\mathrm{cm}}\simeq0.2\,\mathrm{MeV}$, whereas the $\Eone$ rises and takes over the
upper window. 
The lower panel of Fig.~\ref{fig:sigma} compares the posterior with various theoretical determinations~\cite{Rupak:1999rk,Acharya:2021lrv,Du:2022zds}. 
Generally, the agreement is quite good, with the chiral-EFT~\cite{Acharya:2021lrv,Du:2022zds} sitting a few percent lower than our best determination,
and the pionless-EFT~\cite{Rupak:1999rk} agreement is sub-percent except at the very top of the BBN window, where it reaches about $1\%$.

Figure~\ref{fig:sfactor} translates the cross section into $S(E)$ across the BBN band, including the $\pm1\sigma$ posterior.
The one absolute direct-capture datum inside the window, $35.2\pm2.4\,\mu\mathrm{b}$
at $E_{\mathrm{cm}}=275\,\mathrm{keV}$~\cite{Nagai:1997zz}, sits $0.9\sigma$ below
the prediction; neither it nor the near-threshold capture points of
Ref.~\cite{Suzuki1995} enter the fit, so both are genuine cross-checks. 
The band is $0.22\%$ at the low edge, where the magnetic dipole is tied to the sub-percent thermal measurements, dips to $0.19\%$ near $200\,\mathrm{keV}$, and grows to $0.24\%$ at $600\,\mathrm{keV}$, where the $\Eone$ contact's uncertainty dominates; the treatment systematic adds a further $0.05\%$ at $100\,\mathrm{keV}$ and $0.14\%$ at $600\,\mathrm{keV}$, negligible at the low edge.
Averaging over the Maxwell--Boltzmann distribution gives the
thermal rate $N_A\langle\sigma v\rangle(T)$, determined parametrically to $0.21$--$0.24\%$ over
$T_{9}=0.01$--$1$ and $0.20\%$ at $T_{9}=3$, where the
electric contact coefficient leaves more freedom than the magnetic dipole.
The SLEGS-treatment systematic adds up to $0.06\%$ below $T_{9}=1$ and $0.10\%$ at $T_{9}=3$,
and the ${}^{3}P$ one a further $0.02\%$ and $0.03\%$,
for a combined $0.22$--$0.24\%$ over $T_{9}=0.01$--$1$ and $0.23\%$ at $T_{9}=3$. It reproduces the
independent dibaryon-EFT determination of Ref.~\cite{Ando:2005cz} to $0.10\%$ at $T_{9}=0.3$ and $0.40\%$ at $T_{9}=1$.

\begin{figure}[!tb]
  \centering
  \includegraphics[width=\columnwidth]{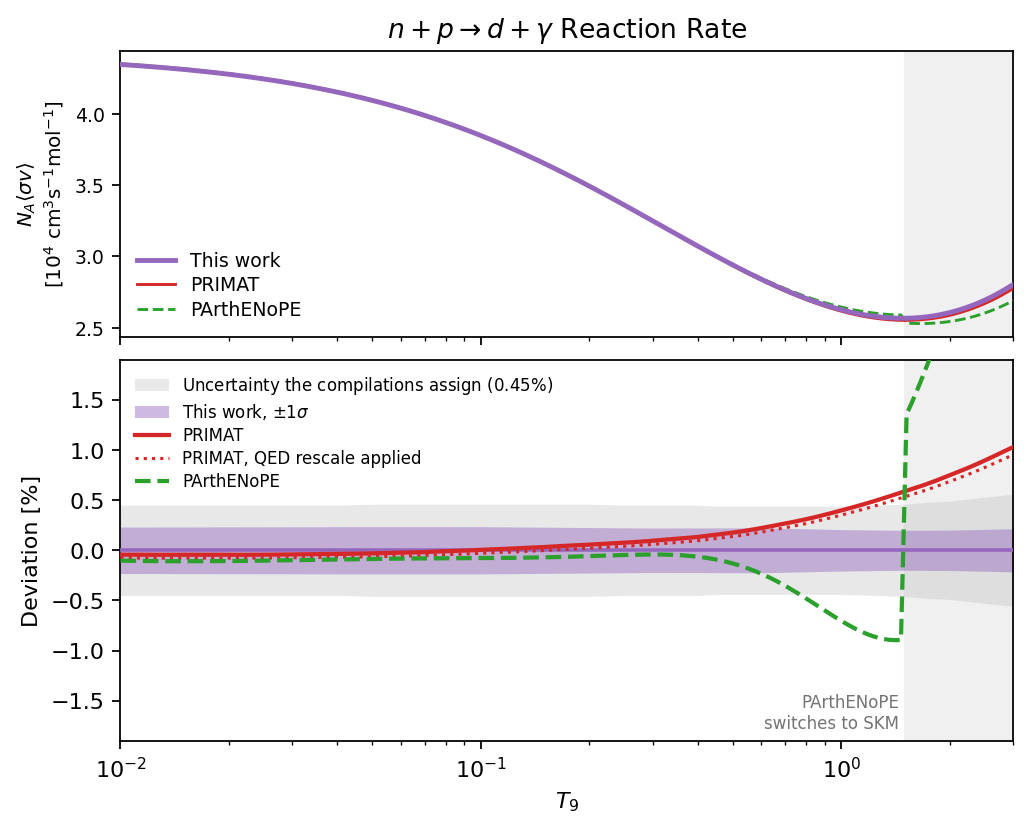}
  \caption{The thermonuclear rate $N_{A}\langle\sigma v\rangle$ (upper) and its
  deviation from the rates implemented by various BBN codes (lower). The shaded region marks
  $T_{9}>1.5$, above which PArthENoPE switches parameterization. Bands are our
  $\pm1\sigma$ posterior and the uncertainty assigned by the compilations.}
  \label{fig:rate}
\end{figure}

Figure~\ref{fig:rate} compares the rate with implementations used by popular BBN codes.
PRIMAT~\cite{Pitrou:2018cgg} tabulates the dibaryon-EFT
result of Ref.~\cite{Ando:2005cz}, and applies a QED pair-production rescale on top by
default.
AlterBBN~\cite{Arbey:2018zfh} also implements that rate, as a rational function of
$T_{9}$ that reproduces the PRIMAT table to $0.003\%$ below
$T_{9}=1.5$, and above it switches to the Smith--Kawano--Malaney
fit~\cite{Smith:1992yy}. PArthENoPE~\cite{Pisanti:2007hk} carries the fit of
Ref.~\cite{Serpico:2004gx}, built on the pionless result of
Ref.~\cite{Rupak:1999rk} below $T_{9}=1.5$, and above it the same
Smith--Kawano--Malaney expression that AlterBBN uses.
Below the nucleosynthesis window all of them agree with our result to
better than $0.1\%$. The differences appear where the $\Eone$ contact does: the
tabulated rate sits $0.40\%$ low at $T_{9}=1$ and $1.0\%$ low at $T_{9}=3$, against
the $0.45\%$ the compilations assign themselves, and the fit of
Ref.~\cite{Serpico:2004gx} is $0.70\%$ high at $T_{9}=1$ before its switch.
AlterBBN raises its own uncertainty from $0.45\%$ to $7.8\%$ across that
switch, which is a fair statement of how well the high-temperature branch is known.
Our band is at or below $0.24\%$ throughout, with no such division.

The uncertainties quoted above are those of the posterior. The truncation of the
expansion is bounded separately and does not enter the likelihood, so it is not
contained in the band just quoted; we state it here and then combine the two.
Three towers are omitted, and one scheme is chosen. The contact terms of the next order --- the
momentum-insertion currents of Sec.~\ref{sec:bdy} --- are degenerate with
$c_{\Mone}$ and $c_{\Eone}$ in the unpolarized cross section: their energy
dependence tracks $\Egam/\Lambda$ rather than $\prel/\Lambda$, and for coefficients
of natural size the part the two fitted constants cannot absorb moves $\sigma(E)$
by $0.03\%$ in quadrature over the seven new directions, nearly flat from threshold
to $1\,\mathrm{MeV}$. Those seven share a single $\Egam^{2}$ shape, so if their
unknown coefficients happened to align the bound would be $0.07\%$ instead. The
effective-range expansion of Eq.~\eqref{eq:xidef} truncates at $v_{2}$, worth less
than $0.03\%$ across the nucleosynthesis window. The corrections dropped in
Sec.~\ref{sec:xsec} --- the deuteron $D$ state and the relativistic terms ---
together move $\sigma(E)$ by less than $0.1\%$. The scheme is the one fixing the
electric contact's own rescattering (Appendix~\ref{app:E1dress}): refitting with
the contact dressed exactly as the impulse term moves $c_{\Eone}$ by $+0.07$ but
$\sigma(E)$ and the rate by less than $0.04\%$. Added in quadrature the four come
to $0.12\%$, a factor of two inside the posterior band: the numbers above remain limited by the data rather than by the
EFT expansion, albeit not by much. Folding the truncation in gives a total theory
uncertainty on the rate of $0.25$--$0.27\%$ across $T_{9}=0.01$--$3$. The rate
and its total uncertainty are tabulated on the PRIMAT temperature grid in the
machine-readable ancillary file \texttt{npdg\_rate\_table.csv} distributed with
this article.

\section{Discussion and Outlook}
\label{sec:discussion}

Our results demonstrate that the on-shell amplitude EFT framework introduced in
Ref.~\cite{Tait:2026guk} can be successfully applied to a second radiative capture. 
The deuteron $d$--$n$--$p$ vertex is matched to the ANC, and the leading low-energy
EFT corrections are captured by the $c_{\Mone}$ and $c_{\Eone}$ coefficients, fixed by the thermal cross section
and photodisintegration data, respectively.   A fit to the experimental data with theory inputs treated as priors provides
an estimate for deuterium production across the nucleosynthesis window, in agreement with
independent pionless and dibaryon-EFT calculations. The predicted
photodisintegration cross section matches the full SLEGS data set. The joint fit delivers the production rate at the
$0.22$--$0.24\%$ level, including the SLEGS- and ${}^{3}P$-treatment systematics, and
at $0.25$--$0.27\%$ once the truncation of the expansion is folded in. That is sufficient, for
practical purposes, to retire this reaction from the primordial deuterium error
budget.

We propagate the posterior rate through a BBN network
(PRyMordial~\cite{Burns:2023sgx}, with all other reactions set to the PRIMAT
compilation~\cite{Pitrou:2018cgg}). The predicted primordial deuterium shifts by
$\Delta(\mathrm{D/H})=-0.06\%$ relative to the standard compilation rate, and
the $\npdg$-induced uncertainty in D/H falls from $0.089\%$ to
$0.050\%$, a factor of $1.8$. That figure propagates the full theory uncertainty of
Sec.~\ref{sec:results}---the $0.25$--$0.27\%$ total, treatment systematics and
truncation included---applied to the rate as a fully correlated shift, matching how
the compilation's own $0.45\%$ is propagated; the posterior band alone would give
$0.044\%$.\footnote{The low-temperature network is
integrated with a relative tolerance of $10^{-6}$; at the solver's default
setting the predicted D/H is converged only at the few$\times10^{-4}$ level,
comparable to the shift quoted here. Tightening further to $10^{-8}$ moves D/H
by $0.002\%$.} The residual is dominated
by the electric-contact uncertainty above the deuterium bottleneck, precisely the
region where the SLEGS beam spectra, rather than their quoted uncertainties, now set
the limit. The treatment
systematic moves the rate by less than $0.04\%$ below $T_{9}\simeq0.6$, which is why
folding the full uncertainty into the network costs only $0.006\%$ in D/H over the
posterior alone, and leaves the central deuterium shift untouched. Of course, the production reaction was already considered 
subdominant in the nuclear error budget for primordial deuterium.  The $d+d$ burning reactions dominate,
consistent with the conclusion the SLEGS Collaboration drew from their own
dibaryon-EFT analysis~\cite{Chen:2025oyj}. The corresponding shift in the baryon
density inferred from D/H is $+0.04\%$, negligible at current precision.

Two extensions are obvious next steps. The next-to-leading-order contact basis, the seven new
parity-even structures with one momentum insertion enumerated in Sec.~\ref{sec:bdy},
feeds the polarization and angular-distribution observables: the
$\Mone$--$\Eone$ interference survives in $d\sigma/d\Omega$, and the $^{3}P_{2}$ and
$d$-wave channels that open at that order carry the first $\Ethree$ and $\Mthree$ contributions.
The initial-state interaction that
the unpolarized cross section sees as a sub-percent correction is a twelvefold
cancellation, so a $J$-resolving observable sees it some thirty times larger. And
Eq.~\eqref{eq:cE1P0} predicts that the leading electric contact sits in $^{3}P_{0}$
alone: the $^{3}P_{2}$ electric dipole that appears one order up is degenerate with
it in the unpolarized cross section, and only a channel-resolved measurement can
separate the two.

It is unsurprising that the results here agree so well with the pionless-EFT
calculation of Ref.~\cite{Rupak:1999rk}: the underlying EFT is practically
identical, and the two are matched to similar experimental inputs.
One advantage of the amplitude program is its assembly of the full amplitude from
on-shell building blocks that are shared among related reactions.  For example,
the $d$--$n$--$p$ vertex also links this production process 
to the $d+d$ reactions, revealing correlations that are less apparent in an EFT formulated in, e.g.,
heavy-particle language.  We plan to complete the deuterium-burning reactions next; a simultaneous fit
to all four primary reactions would then control such correlations directly.

\begin{acknowledgments}

The author is grateful for illuminating conversations with Sonia Bacca, Vincenzo Cirigliano, 
Cara Giovanetti, Shirley Li, Martin Savage, and especially Mauro Valli for suggesting this work as an interesting problem.
The author thanks the SLEGS Collaboration for providing the measured $\gamma$-beam energy distributions and for helpful correspondence about their analysis.
The research of TMPT is supported in part by the National Science Foundation through Grant PHY-2514888
and was supported by API credits provided through the OpenAI Researcher Access Program.
Feynman diagrams were produced with the Feynmangraphz package by Michael Ratz.\\

\noindent
{\bf AI Usage:}~All analytic calculations and numerical results were primarily derived using Claude Opus 4.8/5.0 or Fable 5.0 by Anthropic and GPT-5.6 Sol by OpenAI, 
guided and cross-checked/validated by the human author, who takes full responsibility for all of the results presented here.
\end{acknowledgments}

\appendix

\section{Rescattering}
\label{app:bubble}

This appendix introduces the four-nucleon interaction amplitudes and uses them to compute the initial-state rescattering corrections to the $s$-wave entrance channels.
Rescattering dresses the bare capture process in two important ways.  Summing the ladder of bubble diagrams (see Fig.~\ref{fig:bubblechain}) produces
an elastic multiplicative factor that depends on the multipole channel, and the exit from the ladder effectively converts the tree-level deuterium production amplitudes into loop diagrams,
dressing their ingredients in a way that differentiates the factorized and boundary terms.

The physical picture is that the deuterium production is short-ranged, probing the $np$ pair at short distances.  Resumming the rescattering ladder
converts the incoming plane waves into the full $^{1}S_{0}$ scattering state---the Watson enhancement, which is large because of the virtual
state sitting just below threshold. In the crossed photodisintegration process, the
same analytic function describes the final-state interactions of the outgoing nucleon pair, allowing the SLEGS data to precisely test the same matrix elements.

\subsection{Four-Nucleon Amplitudes and Couplings}
\label{app:nnamps}

In the EFT language, the elastic $np$ amplitude has a factorized piece from deuteron exchange --- which only contributes to the ${}^{3}S_1$ partial wave ---  as well as boundary terms: four-nucleon contact interactions.  The deuteron exchange is already computable in terms of the strong vertex of Sec.~\ref{sec:strongvertex}.  For the contact terms, with all legs incoming,
$(1,2,3,4)=(n,\bar n,p,\bar p)$, and $s_{np}=s_{13}$, the basis of lowest dimension parity-even amplitudes is
\begin{equation}
\begin{split}
  \mathcal{M}^{(0)}_{4N} &= \frac{1}{\Lambda^{2}}\sum_{k=1}^{5} c_k\,\mathcal{P}_k,\\
  \mathcal{P}_1 &= \ad{\bm1}{\bm2}\ad{\bm3}{\bm4}
                 + \sq{\bm1}{\bm2}\sq{\bm3}{\bm4},\\
  \mathcal{P}_2 &= \ad{\bm1}{\bm4}\ad{\bm2}{\bm3}
                 + \sq{\bm1}{\bm4}\sq{\bm2}{\bm3},\\
  \mathcal{P}_3 &= \ad{\bm1}{\bm2}\sq{\bm3}{\bm4}
                 + \sq{\bm1}{\bm2}\ad{\bm3}{\bm4},\\
  \mathcal{P}_4 &= \ad{\bm1}{\bm3}\sq{\bm2}{\bm4}
                 + \sq{\bm1}{\bm3}\ad{\bm2}{\bm4},\\
  \mathcal{P}_5 &= \ad{\bm1}{\bm4}\sq{\bm2}{\bm3}
                 + \sq{\bm1}{\bm4}\ad{\bm2}{\bm3},
\end{split}
\label{eq:nnbasis}
\end{equation}
all five of which are independent. The four-nucleon operators are of
dimension six, so the explicit $1/\Lambda^{2}$ leaves the $c_k$---and the
channel couplings built from them below---dimensionless, on the same footing as
the electromagnetic contacts of Sec.~\ref{sec:xsec}.

Projected onto two-nucleon channels, the $s$-wave couplings descend from two combinations of the $c_k$,
\begin{align}
  g_A &= c_3+c_5, \nonumber\\
  g_B &= c_1+c_2+2c_4, \nonumber\\
  h_A &= c_5-c_3, \nonumber\\
  h_B &= c_2-c_1,
  \label{eq:gAB}
\end{align}
where $g_{A,B}$ correspond to $^{1}S_{0}$ and $h_{A,B}$ to $^{3}S_{1}$.
The remaining combination, $\mathcal{P}_1+\mathcal{P}_2-\mathcal{P}_4$, feeds the spin-triplet $P$-waves, contributing at relative order $\prel^{2}/m_N^{2}$.
Writing $C_\alpha(s_{np})$ for the projected contact amplitude of channel
$\alpha$ as a function of the $np$ invariant mass $s_{np}=(p_n+p_p)^{2}$, the
two projections are
\begin{align}
  C_{^1S_0}(s_{np}) &= \frac{1}{\Lambda^{2}}
      \left[g_A + g_B\,\frac{s_{np}}{4m_N^{2}}\right],
  \label{eq:Csinglet}\\
  C_{^3S_1}(s_{np}) &= \frac{1}{\Lambda^{2}}\,
      \frac{h_B(E+2m_N)^{2}+h_A(2E+m_N)^{2}}{18m_N^{2}},
  \label{eq:Ctriplet}
\end{align}
with $E=\tfrac{1}{2}\sqrt{s_{np}}$. The singlet is linear in $s_{np}$, the
triplet analytic through the fermion factor $E$. At threshold, $s_{np}=4m_N^{2}$, they take the values $(g_A+g_B)/\Lambda^{2}$ and
$(h_A+h_B)/(2\Lambda^{2})$; the leading slope in $s_{np}$ furnishes the recoil-locked
correction, and the nonrelativistic expansion in $\prel^{2}=s_{np}/4-m_N^{2}$
is Eq.~\eqref{eq:NRform}.
The $^{3}S_{1}$--$^{3}D_{1}$ mixing factorizes as a rank-one coupled-channel element of relative size $\prel^{2}/6m_N^{2}$ at threshold. 
Its size, $10^{-4}$ across the nucleosynthesis window and below $10^{-3}$ at the top of the photodisintegration range, places it far below our resolution.

At higher EFT orders, the $^{1}S_{0}$ projection leads to no new energy-dependence at dimension seven (just corrections to existing forms),
and to a single new shape at dimension eight proportional to $s_{np}^{2}$, parameterized by the $^{1}S_{0}$ shape parameter $v_{2}$.
The $^{3}S_{1}$ operators contain a new energy-dependent form at dimension seven, so its shape parameter $v_{2}^{t}$ arises one order before $v_{2}$.
Finally, the $^{3}P_{J}$ channels receive corrections at dimension eight which include contributions to the structure already present at dimension six, but are significant
because they are order $\prel^{2}/\Lambda^{2}$, and thus naturally dominate the contribution from the $\prel^{2}/m_N^{2}$-suppressed dimension six term.

\subsection{Rescattering and Resummation}

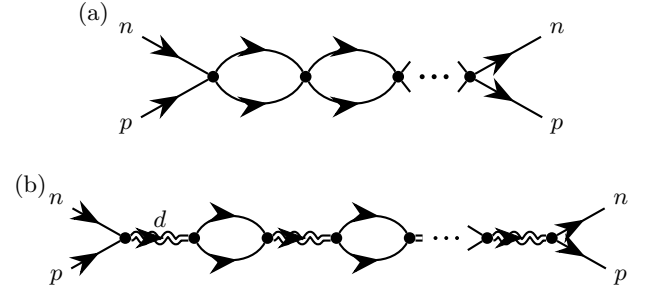
\begin{figure}[t]
\centering
\begin{tikzpicture}[fmg, every node/.append style={font=\small}, scale=0.72]
  \node at (-2.2,1.15) {(a)};
  \path (-1.6,0.9) node (n) {$n$} (-1.6,-0.9) node (p) {$p$}
        (0,0) node[vertex,name=C1] {};
  \draw[nucl] (n) -- (C1); \draw[nucl] (p) -- (C1);
  \path (1.7,0) node[vertex,name=C2] {};
  \draw[nucl] (C1) to[bend left=60] (C2);
  \draw[nucl] (C1) to[bend right=60] (C2);
  \path (3.4,0) node[vertex,name=C3] {};
  \draw[nucl] (C2) to[bend left=60] (C3);
  \draw[nucl] (C2) to[bend right=60] (C3);
  \draw[thick] (C3) -- (3.62,0.28); \draw[thick] (C3) -- (3.62,-0.28);
  \foreach \x in {3.84,4.06,4.28} \fill (\x,0) circle (1.3pt);
  \path (4.72,0) node[vertex,name=C4] {};
  \draw[thick] (4.50,0.28) -- (C4); \draw[thick] (4.50,-0.28) -- (C4);
  \path (6.32,0.9) node (nout) {$n$} (6.32,-0.9) node (pout) {$p$};
  \draw[nucl] (C4) -- (nout); \draw[nucl] (C4) -- (pout);
\end{tikzpicture}\\[0.9em]
\begin{tikzpicture}[fmg, every node/.append style={font=\small}, scale=0.57]
  \node at (-2.2,1.2) {(b)};
  \path (-1.6,0.9) node (n) {$n$} (-1.6,-0.9) node (p) {$p$}
        (0,0) node[vertex,name=V1] {};
  \draw[nucl] (n) -- (V1); \draw[nucl] (p) -- (V1);
  \path (1.6,0) node[vertex,name=V2] {};
  \dwave{V1}{V2}
  \path (0.8,0.45) node {$d$};
  \path (3.3,0) node[vertex,name=V3] {};
  \draw[nucl] (V2) to[bend left=60] (V3);
  \draw[nucl] (V2) to[bend right=60] (V3);
  \path (4.9,0) node[vertex,name=V4] {};
  \dwave{V3}{V4}
  \path (6.6,0) node[vertex,name=V5] {};
  \draw[nucl] (V4) to[bend left=60] (V5);
  \draw[nucl] (V4) to[bend right=60] (V5);
  \draw[thick,double,double distance=1.1pt] (V5) -- (6.9,0);
  \foreach \x in {7.2,7.45,7.7} \fill (\x,0) circle (1.3pt);
  \path (8.4,0) node[vertex,name=V6] {};
  \draw[thick] (7.95,0.28) -- (V6); \draw[thick] (7.95,-0.28) -- (V6);
  \path (9.9,0) node[vertex,name=V7] {};
  \dwave{V6}{V7}
  \path (11.5,0.9) node (nout) {$n$} (11.5,-0.9) node (pout) {$p$};
  \draw[nucl] (V7) -- (nout); \draw[nucl] (V7) -- (pout);
\end{tikzpicture}
\caption{The elastic rescattering ladder resummed in Eq.~\eqref{eq:geosum}, including
(a) contact interactions $C(s_{np})$, and (b) deuteron exchange dressed by nucleon bubble insertions ($^{3}S_{1}$ only).}
\label{fig:ladders}
\end{figure}

Schematically, resumming the ladder of $s$-wave bubble diagrams produces the factor $R(s_{np})$,
\begin{equation}
\begin{split}
  R(s_{np}) &= 1 + 4m_N^{2}\,V\Pi + \big(4m_N^{2}\,V\Pi\big)^{2} + \cdots\\
            &= \frac{1}{1 - 4m_N^{2}\,V(s_{np})\,\Pi(s_{np})},
\end{split}
  \label{eq:geosum}
\end{equation}
where $V(s_{np})$ is one insertion of the tree-level $n p \rightarrow n p$ scattering amplitude (Fig.~\ref{fig:ladders}):
\begin{equation}
  V(s_{np}) =
  \begin{cases}
    C_{^1S_0}(s_{np}) & (^1S_0)\\[6pt]
    \dfrac{\CSv^{2}}{\,s_{np}-m_{d}^{2}\,} + C_{^3S_1}(s_{np}) & (^3S_1)
  \end{cases}
  \label{eq:Vdef}
\end{equation}
where the deuteron exchange enters only the ${}^{3}S_1$ channel, with $\CSv$
the $dnp$ coupling of Eq.~\eqref{eq:strongSD} and $m_{d}$ the deuteron mass.
Both are written throughout as the renormalized, on-shell quantities they become
below. Here $C(s_{np})$ is the $n$-$p$-$n$-$p$ vertex of
Eqs.~\eqref{eq:Csinglet}--\eqref{eq:Ctriplet} projected onto a given channel,
$\Pi(s_{np})$ is the contribution from a single nucleon bubble, and the $4m_N^{2}$ is the
relativistic normalization of the two nucleon legs.

A single bubble is assembled from the same on-shell amplitudes, through its
discontinuity. Cutting the loop places the two intermediate nucleons on shell,
with the $V$ on either side; the
discontinuity is the two-body phase space, built from on-shell amplitudes,
\begin{equation}
  \mathrm{Im}\,\Pi(s_{np}) = \frac{\beta}{16\pi},
  \qquad \beta = \sqrt{1 - \frac{4m_N^{2}}{s_{np}}}\,.
  \label{eq:ImPi}
\end{equation}
The bubble grows only logarithmically, so a once-subtracted dispersion relation
reconstructs it from this cut,
\begin{equation}
\begin{split}
  \Pi(s_{np}) &= \Pi(0) + \frac{s_{np}}{\pi}
     \int_{4m_N^{2}}^{\infty}\!\frac{\mathrm{Im}\,\Pi(s')}{s'\,(s'-s_{np})}\,ds'\\
   &= \frac{B_0(s_{np};\,m_N,m_N)}{16\pi^{2}},
\end{split}
  \label{eq:PiB0}
\end{equation}
with $B_0$ the scalar two-point function~\cite{tHooft:1978xw},
\begin{equation}
  B_0(s_{np}) = \Delta_\epsilon + 2 - \ln\frac{m_N^{2}}{\mu^{2}}
   - \beta\Big[\ln\frac{1+\beta}{1-\beta} - i\pi\Big],
  \label{eq:B0}
\end{equation}
where $\Delta_\epsilon$ is the $\overline{\mathrm{MS}}$ pole and $\mu$ the renormalization scale. The
dispersion relation supplies more than the cut: the subtraction constant
$\Pi(0)$ and the analytic part of the integral are precisely the real terms a
covariant loop would carry as its divergence $\Delta_\epsilon$, its scale $\mu$,
and the even series $-\beta\ln[(1+\beta)/(1-\beta)] = -2\beta^{2} -
\tfrac23\beta^{4} - \cdots$.  The one nonanalytic piece, the $i\pi\beta$
discontinuity, continues below threshold to $-\pi w(s_{np})$ with
\begin{equation}
  w(s_{np}) \equiv \sqrt{\frac{4m_N^{2}}{s_{np}} - 1}
   = \frac{2}{\sqrt{s_{np}}}\,\sqrt{m_N^{2} - \frac{s_{np}}{4}}\,,
  \label{eq:wdef}
\end{equation}
real below threshold, and $w \to -i\beta$ above it.

\subsubsection{${}^{1}S_0$ Channel}

The infinities (and finite corrections) in $\Pi$ renormalize the coefficients in the tower of the $n$-$p$-$n$-$p$ interactions.
In terms of $T \equiv V R$, inverting the geometric sum of Eq.~\eqref{eq:geosum},
\begin{equation}
\begin{split}
  \frac{1}{T_{^1S_0}(s_{np})} &= \frac{1}{C_{^1S_0}(s_{np})} - 4m_N^{2}\,\Pi(s_{np})\\
  &\rightarrow \frac{m_N}{4\pi}\,\xi(\prel^2) + \frac{m_N^{2}\,w(s_{np})}{4\pi},
\end{split}
\label{eq:Minv}
\end{equation}
where $\xi(\prel^2)$ is the renormalized polynomial of Eq.~\eqref{eq:xidef}.
The analytic content of $\Pi$, including the $\log$-divergence---everything but
the $w$ cut---is real and polynomial in $(s_{np}-4m_N^{2})$ near threshold, and
can be absorbed into $1/C(s_{np})$, renormalizing the EFT tower of $n$-$p$-$n$-$p$
couplings, order-by-order.

With $s_{np} = (2m_N + E)^2$, the branch factor becomes
\begin{equation}
  \sqrt{m_N^2 - \tfrac{s_{np}}{4}} = \kappa(E)\Big(1 + \tfrac{E}{8m_N} + \dots\Big),
  \quad \kappa(E) \equiv \sqrt{-m_N E},
\end{equation}
with $\kappa>0$ below threshold and $\kappa \to -i\prel$ on the cut. Collecting
the even-power relativistic corrections into the polynomial\footnote{The $\prel$ odd-power relativistic remnants cannot be absorbed: the cut term is $-i\prel\,[1 - 3\prel^2/(8m_N^2) + \dots]$. 
They represent a $2\times10^{-3}$ correction at the top of the SLEGS range and $3\times10^{-4}$ in the BBN window, and are thus negligible.},
\begin{equation}
  \frac{1}{T_{^1S_0}(E)} = \frac{m_N}{4\pi}\Big[\,\xi(\prel^2) + \kappa(E)\,\Big],
\label{eq:NRform}
\end{equation}
where
\begin{equation}
  \xi(\prel^2) \equiv \frac{4\pi}{m_N\,C_{^1S_0}}
      \simeq -\frac{1}{a_0} + \frac{r_0}{2}\prel^2 + v_2 \prel^4 + \dots.
\label{eq:xidef}
\end{equation}
The resulting polynomial $\xi(\prel^2)$ defines the effective-range expansion (ERE) in terms of the scattering length $a_0$ and effective range $r_0$:
\begin{equation}
  -\frac{1}{a_0} = \frac{4\pi}{m_N\,C(0)},\qquad
  r_0 = -\frac{8\pi}{m_N}\,\frac{C'(0)}{C(0)^{2}},
  \label{eq:EREmap}
\end{equation}
which in terms of the renormalized on-shell couplings of Eq.~\eqref{eq:Csinglet} give for the $^{1}S_{0}$ channel,
\begin{align}
  -\frac{1}{a_0} &= \frac{4\pi\Lambda^{2}}{m_N(g_A+g_B)}, &
  r_0 &= -\frac{8\pi\Lambda^{2}}{m_N^{3}}\,\frac{g_B}{(g_A+g_B)^{2}}.
  \label{eq:EREsinglet}
\end{align}
The next term in $\xi$, the shape parameter $v_2\sim1/\Lambda^{3}$, enters at relative order
$(\prel/\Lambda)^{2}$. Its effect across the BBN window is
sub-percent, and truncating at $(a_0,r_0)$ suffices.

\subsubsection{${}^{3}S_1$ Channel}

In the ${}^{3}S_1$ channel the kernel of Eq.~\eqref{eq:Vdef} contains the
deuteron, so the geometric sum dresses its propagator alongside the contact
tower. Attached to the deuteron line, the nucleon bubble of
Eq.~\eqref{eq:PiB0} is the deuteron self-energy,
\begin{equation}
  \Sigma(s_{np}) = \CSv^{2}\,4m_{N}^{2}\,\Pi(s_{np})
  \;\longrightarrow\; -\frac{\CSv^{2}m_{N}}{4\pi}\,\kappa(s_{np}),
  \label{eq:Sigren}
\end{equation}
the arrow indicating that its analytic part has been absorbed: the divergence
of Eq.~\eqref{eq:B0} is $s$-independent and renormalizes the deuteron mass, and
the quadratic and higher terms of the even $\beta$-series, which a single pole
cannot absorb, pass into the contact tower $C_{^3S_1}$. What survives is the
cut, $\kappa=\tfrac12\sqrt{4m_{N}^{2}-s_{np}}$, with $\kappa\to-i\prel$ above
threshold as in Eq.~\eqref{eq:NRform}.

We renormalize on shell, fixing the pole at the measured deuteron mass and the
residue at the measured ANC. The renormalized parameters are then the ones the
capture amplitude already uses, $m_{d}$ and $\CSv$.

That residue is related to the quantity $\rho_{d}$, defined by normalizing the deuteron radial
wavefunction so that $u(r)\to e^{-\gamma r}$ outside the range,
\begin{equation}
  \rho_{d} = 2\int_{0}^{\infty}\!\!dr\,\Big[e^{-2\gamma r}-u(r)^{2}\Big]
           = \frac{1}{\gamma}-\frac{2}{\AS^{2}}\,.
  \label{eq:rhoint}
\end{equation}
$\rho_{d}$ measures the deficit of the true wavefunction against its own
asymptotic tail---the norm the interior fails to supply relative to
extrapolating $e^{-\gamma r}$ all the way in. It is a length,
$1.76\,\mathrm{fm}$, of order $1/m_\pi$ as expected. 
A pure exponential of the same binding would have $\AS^{2}=2\gamma$;
because the real deuteron falls short inside, its tail must be larger by
$1/(1-\gamma\rho_{d})=1.69$ to carry the norm. In quantum field theory language,
$\rho_d$ is related to the wavefunction renormalization
$\mathcal{Z}_{d}=-\gamma\rho_{d}/(1-\gamma\rho_{d})$.\footnote{$\mathcal{Z}_{d}=-0.69$. An elementary field would have
$0<\mathcal{Z}\le1$; a negative norm is the signature of an auxiliary field
standing in for a composite state---the same fact that requires a wrong-sign
kinetic term in the dibaryon formulation~\cite{Kaplan:1996nv}, and that
Weinberg's compositeness argument reads off the sign of the effective
range~\cite{Weinberg:1965zz}.}
Matching the residue to the on-shell vertex gives
\begin{equation}
  \CSv = \sqrt{\frac{8\pi\gamma}{m_{N}\,(1-\gamma\rho_{d})}} \;=\; 1.438 ,
  \label{eq:CSgd}
\end{equation}
to be compared with $\AS\sqrt{2\pi m_{d}}/m_{N}=1.437$ from the measured ANC. 

Expanding the same construction about threshold instead of about the pole
returns the effective-range expansion. With $\xi_t\equiv4\pi/(m_N V)$,
\begin{equation}
  \frac{1}{T_{^3S_1}(E)} = \frac{m_N}{4\pi}\big[\,\xi_t(\prel^2)+\kappa(E)\,\big],
\end{equation}
and $\rho_{d}=2\,\xi_t'(-\gamma^{2})$ while $r_{t}=2\,\xi_t'(0)$. 
Deuteron exchange alone gives
$\xi_t$ linear in $\prel^{2}$, so by itself it predicts $\rho_{d}=r_{t}$ and
$v_2^t=0$, with every departure due to the tower of contact interactions
$C_{^3S_1}(s_{np})$ and its higher-dimension partners. Because the deuteron is
an external state of both retained multipoles, the ${}^{3}S_1$ amplitude enters
them only at its pole, never at the reaction energy: the momentum governing
those corrections is the binding $\gamma$, not the small
$\prel$.  So whereas $v_2$ is safely dropped, $v_2^t$ is retained by extracting $\rho_{d}$ at the pole
(rather than the threshold), which resums it and every higher $v_n^t$ at once.

\subsubsection{${}^{3}P_J$ Channels}

The situation is quite different for the $\Eone$ $p$-wave initial states.
Eq.~\eqref{eq:NRform} applies to any partial wave, and for $\ell=1$ the centrifugal barrier puts three powers of momentum on the cut,
\begin{equation}
  \frac{1}{T_{^{3}P_J}(E)} = \frac{m_N}{4\pi}\Big[
  -\frac{1}{a_{1}^{(J)}} + \mathcal{O}(\prel^{2}) - i\,\prel^{3}\Big],
  \label{eq:Pwave}
\end{equation}
with $a_{1}^{(J)}$ the scattering volume. Each rescattering is a factor
$\tan\delta_{^{3}P_J}\simeq -\,a_{1}^{(J)}\prel^{3}$, with $|a_{1}|\sim1/m_\pi^{3}$ since there are no (real or virtual) poles
near threshold.  The interaction is a contact, so the geometric series sums into a
single insertion of the full elastic amplitude, exactly as it does in the $^{1}S_{0}$
channel above. Closing that insertion onto the tree amplitude
(Appendix~\ref{app:E1dress}) gives the initial-state correction
$\delta_{\Eone}^{(J)}$ of Sec.~\ref{sec:isi}. We keep the full amplitude rather
than its threshold limit. Truncating it there would be safe channel by channel,
since $1/|a_{1}|\gg\prel^{3}$ throughout, but the observable depends on the
$(2J+1)/9$ average of the three channels, and that average is a twelvefold
cancellation whose balance shifts with energy: the individual
$\delta_{\Eone}^{(J)}$ reach tens of percent at the top of the fitted range where
their average is below one.

\subsection{Matching to Scattering Data}
\label{app:match}

The matrix elements $T$ in each channel describe elastic scattering of nucleons, and measurements of this process can thus be
used to fit the undetermined $n$-$p$-$n$-$p$ scattering coefficients from which they are built.
The Nijmegen partial-wave analysis (PWA93)~\cite{Stoks:1993tb}
is an energy-dependent fit to the $np$ scattering database, and the compilation of
Ref.~\cite{deSwart:1995ui} reports its ${}^{3}S_{1}$-and-deuteron
content---the triplet scattering length, effective range, and shape parameters through
$v_5$---as threshold derivatives. 

We fit the low-energy constants of the singlet channel to the PWA93 phases. In that case,
$\xi=4\pi/(m_N C_{^1S_0})$ has no left-hand singularity nearer than the one-pion
branch point at $\prel^2=-m_\pi^2/4$, so its two-term expansion converges for
$|\prel|<m_\pi/2$. Fitting the phases inside that radius gives
\begin{equation}
  a_0^{\mathrm{exp}} = -23.61(8)\,\mathrm{fm}, \qquad
  r_0^{\mathrm{exp}} = 2.630(7)\,\mathrm{fm},
  \label{eq:a0r0exp}
\end{equation}
and through Eq.~\eqref{eq:EREsinglet} the scattering length fixes the leading
combination $g_A+g_B$ directly, whereas the effective range isolates $g_B$.

The uncertainties quoted in Eq.~\eqref{eq:a0r0exp} are only statistical, and systematics are expected to be important. 
The $np$ ${}^{1}S_0$ effective range is spread across partial-wave
analyses, $r_0=2.63$--$2.77\,\mathrm{fm}$ according to the treatment of
electromagnetic corrections and the energy weighting, and our window fit sits at
the low edge of that family. 
The joint fit of Sec.~\ref{sec:jointfit} centers the prior on $r_0$ at the family value with the
family spread as its width,  which covers Eq.~\eqref{eq:a0r0exp} at $1.3\sigma$.

\subsection{Dressed Amplitude}
\label{app:raddress}

\begin{figure}[t]
\centering
\begin{tikzpicture}[fmg, every node/.append style={font=\small}, scale=0.72]
  \node at (-0.65,1.45) {(a)};
  \path (-0.45,1.00) node (n) {$n$} (-0.45,-1.00) node (p) {$p$}
        (1.35,0) node[vertex,minimum size=1.1em,name=T] {}
        (4.35,0) node[blob,minimum size=2.4em,name=K] {}
        (6.95,1.20) node (d) {$d$} (6.95,-1.20) node (g) {$\gamma$};
  \draw[nucl] (n) -- (T); \draw[nucl] (p) -- (T);
  \draw[nucl] (T) to[bend left=52] (K);
  \draw[nucl] (T) to[bend right=52] (K);
  \dwave{K}{d} \draw[phot] (K) -- (g);
  \cutint{2.30}{1.55}
  \node at (1.35,-1.85) {$T$};
  \node at (4.55,-1.85) {$\mathcal{M}^{\mathrm{ct}}$};
\end{tikzpicture}\\[0.9em]
\begin{tikzpicture}[fmg, every node/.append style={font=\small}, scale=0.72]
  \node at (-0.65,1.45) {(b)};
  \path (-0.45,1.00) node (n) {$n$} (-0.45,-1.00) node (p) {$p$}
        (1.35,0) node[vertex,minimum size=1.1em,name=T] {}
        (4.45,1.10) node[vertex,name=S] {}
        (4.45,-1.10) node[vertex,name=E] {}
        (6.95,1.10) node (d) {$d$} (6.95,-1.10) node (g) {$\gamma$};
  \draw[nucl] (n) -- (T); \draw[nucl] (p) -- (T);
  \draw[nucl] (T) -- (S); \draw[nucl] (T) -- (E);
  \draw[nucl] (E) -- (S);
  \dwave{S}{d} \draw[phot] (E) -- (g);
  \cutint{2.40}{1.50}
  \node[left=1pt] at (4.45,0) {$p$};
  \node at (1.35,-1.95) {$T$};
  \node at (5.30,-1.95) {$\mathcal{M}^{\mathrm{fact}}$};
\end{tikzpicture}
\caption{The elastic $np$ amplitude $T$ glued across the two-nucleon cut
(dashed) to (a) the boundary piece and (b) the factorized piece of the
tree $np\to d\gamma$ amplitude.}
\label{fig:cutdressing}
\end{figure}
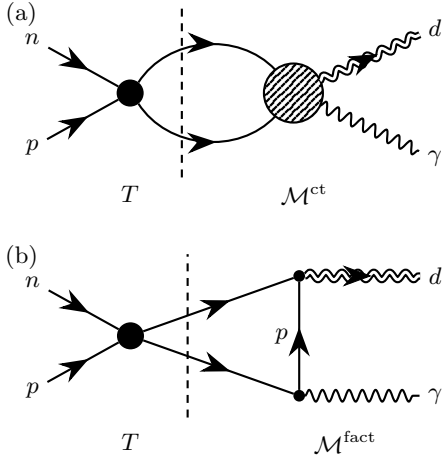

At the end of the rescattering ladder, the final nucleon pair closes onto the tree-level production amplitude, dressing it into a loop diagram
(Fig.~\ref{fig:cutdressing}).
The dispersion techniques that computed the nucleon bubble extract the dressed form, $\mathcal{M}_{D}$.
Resumming the elastic rescattering ladder supplies the universal denominator, Eq.~\eqref{eq:NRform},
\begin{equation}
  \mathcal{M}_{D}(E) \;=\;
  \frac{\mathcal{N}(E)}{\xi(\prel^{2})+\kappa(E)}\,,
  \label{eq:Rdef}
\end{equation}
with $\mathcal{N}$ the dressed production amplitude.

$\mathcal{M}_{D}$ is the rescattering ladder (resumming to $T(s_{np})$) with its final rung replaced by the $^{1}S_{0}$-projected
tree production amplitude $\mathcal{M}$:
\begin{equation}
\begin{split}
  \mathcal{M}_{D}(s_{np}) &= \mathcal{M}(s_{np})
    - T(s_{np})\,L(s_{np}),\\
  \mathrm{Im}\,L(s_{np}) &= -\,\frac{\beta}{16\pi}\;4m_{N}^{2}\,\mathcal{M}(s_{np}),
\end{split}
  \label{eq:Tcap}
\end{equation}
where $L$ is the loop carrying the two nucleons from their last rescattering
into the tree amplitude; cutting it puts them on shell and leaves the
on-shell tree-level amplitude $\mathcal{M}$.  

The factorized and boundary pieces of $\mathcal{M}$ behave differently
far up the cut, so we treat the pieces of
$\mathcal{N} = \mathcal{N}^{\mathrm{ct}} + \mathcal{N}^{\mathrm{fact}}$ separately below.

\subsubsection{Boundary Term}

The boundary piece of the tree amplitude [Fig.~\ref{fig:diagrams}(d)]
has no singularities in $s_{np}$: after the spin contraction its
$^{1}S_{0}$ component is an analytic function $\lambda(s_{np})$. At tree
level $\lambda \propto (c_{\Mone}/\Lambda^{3})\,\Egam$ with
$\Egam(s_{np}) = (s_{np}-m_{d}^{2})/(2\sqrt{s_{np}})$, but the
dispersion analysis requires nothing beyond its analyticity. 
Gluing the elastic amplitude to it
results in the loop of Fig.~\ref{fig:cutdressing}(a).
Eq.~\eqref{eq:Tcap} evaluates it as
\begin{equation}
  L_{\mathrm{ct}}(s_{np})
  = -\,\lambda(s_{np})\;4m_{N}^{2}\,\Pi(s_{np}),
  \label{eq:Lct}
\end{equation}
proportional to the bubble computed in \eqref{eq:B0}.  Its
cut grows like $\sqrt{s_{np}}$, and requires one subtraction per power
of $s_{np}$, with the infinity $\Delta_{\epsilon}$ absorbed into $c_{\Mone}$,
\begin{equation}
  \mathcal{N}^{\mathrm{ct}}(s_{np})
  = m_{N}\,\Egam(s_{np})\,\ell_{1} + \cdots
  \label{eq:Kct}
\end{equation}
vanishing as $\Egam\to0$, as the soft limit requires of a term with no bremsstrahlung pole~\cite{Low:1958sn},
and $\ell_{1}=2m_{d}W(0)\,c_{\Mone}/\Lambda^{3}$ is the coefficient of
Eq.~\eqref{eq:cM1def}, with $W(0)$ the threshold value of Eq.~\eqref{eq:Wdef}.
Expanded nonrelativistically,
$m_{N}\Egam\to\gamma^{2}+\prel^{2}$, and the dressed contact term contributes
$(\gamma^{2}+\prel^{2})\,\ell_{1}$ to Eq.~\eqref{eq:AM1}.

\subsubsection{Factorized Term}

The factorized term in ${\cal M}$ closes into a triangle [Fig.~\ref{fig:cutdressing}(b)]. 
The magnetic vertex contains one power of $\Egam$ from the field
strength whereas the deuteron vertex is constant.  Contracting the nucleon
spins on the cut results in:
\begin{equation}
  \mathrm{Im}\,L(s_{np})
  = A\,(2m_{N})^{3}\,\frac{1}{32\pi\sqrt{s_{np}}}\,
    \ln\frac{1+\beta}{1-\beta}\,,
  \label{eq:ImCrel}
\end{equation}
where $\beta$ is the velocity of Eq.~\eqref{eq:B0}, and the factor
$A \equiv (1+\kappaV)\,\CSv/(m_{N}\sqrt{2m_{d}}\,)$. 
In deriving this expression, the phase space $\beta$ and photon energy from the magnetic interaction
cancel against the transfer propagator denominator. 
A triangle admits three cuts, but the photon and deuteron legs are pinned to $q_{\gamma}^{2}=0$
and $p_{d}^{2}=m_{d}^{2}$, and only the $s_{np}$ cut is open.

The analysis turns on how the sewn discontinuity behaves
far up the cut. Beyond its leading constant, the on-shell amplitude
carries recoil corrections,
$\mathcal{M}^{\mathrm{fact}}(\prel')
= -A\,[1+\mathcal{O}(\prel'^{2}/m_{N}^{2})]$,
and each power converts the falling discontinuity into a growing one.
Their dispersive reconstruction costs subtractions, contributing to the renormalization of the contact interaction $c_{\Mone}$ (and its higher order EFT companions). The
falling piece remains finite.  Carrying out the dispersion integral produces the standard scalar triangle loop function~\cite{tHooft:1978xw}:
\begin{equation}
\begin{split}
  L(s_{np}) &= A\,(2m_{N})^{3}\,\Egam(s_{np})\,\mathcal{C}(s_{np}),\\
  \mathcal{C}(s_{np})
  &= \frac{1}{\pi}\int_{4m_{N}^{2}}^{\infty}\!\!ds'\,
     \frac{\mathrm{Im}\,L(s')/\big[A(2m_{N})^{3}\Egam(s')\big]}{s'-s_{np}}\\
  &= -\,\frac{C_{0}\big(s_{np},0,m_{d}^{2};m_{N},m_{N},m_{N}\big)}
             {16\pi^{2}}\,,
\end{split}
  \label{eq:Cdisp}
\end{equation}
manifestly finite.

The threshold limit is easily taken from the Feynman-parameter representation of $C_0$, for which the denominator rearranges,
\begin{equation}
\begin{split}
  \mathcal{C} &= \frac{1}{16\pi^{2}}\!\int_{0}^{1}\!
  \frac{dx_{1}\,dx_{2}\,dx_{3}\;\delta(1-x_{1}-x_{2}-x_{3})}{\Delta}\,,\\
  \Delta &= m_{N}^{2}-x_{1}x_{2}\,s_{np}-x_{1}x_{3}\,m_{d}^{2}\\
         &= m_{N}^{2}(1-2u)^{2}
           + 4(1-u)\big(\kappa^{2}x_{2}+\gamma^{2}x_{3}\big),
\end{split}
  \label{eq:Cparam}
\end{equation}
with $u=x_{2}+x_{3}$ and $\kappa^{2}=(4m_{N}^{2}-s_{np})/4$, the $\kappa$ of
Eq.~\eqref{eq:Sigren}. The relativistic binding variable
$\gamma^{2}=(4m_{N}^{2}-m_{d}^{2})/4=m_{N}B-B^{2}/4$ agrees with the $\gamma$
used elsewhere up to a relative $B/4m_{N}\sim10^{-3}$. Below threshold both terms are positive, and for
$\kappa,\gamma\ll m_{N}$ the integral is dominated by the strip around
$u=\tfrac12$, where the $m_{N}^{2}$ term is suppressed and the
denominator falls to $\mathcal{O}(\gamma^{2})$: there all three nucleons
are simultaneously near their mass shells with momenta of order
$\gamma$, the nonrelativistic region, and the enhancement
$m_{N}^{2}/\gamma^{2}$ over a width $|1-2u|\sim\gamma/m_{N}$ outweighs
the rest of the simplex by $m_{N}/\gamma$. The $u$ integration across
the strip is a Lorentzian, and with $x_{2}=uw$,
\begin{equation}
\begin{split}
  \mathcal{C} \;&\longrightarrow\;
  \frac{1}{64\pi m_{N}}\!\int_{0}^{1}\!
  \frac{dw}{\sqrt{\gamma^{2}+(\kappa^{2}-\gamma^{2})\,w}}\\
  &= \frac{1}{32\pi\,m_{N}\,\big[\gamma+\kappa(E)\big]}\,.
\end{split}
  \label{eq:Cnr}
\end{equation}
such that
\begin{equation}
  L(E) = \frac{A\,m_{N}^{2}\,\Egam}{4\pi\,\big[\gamma+\kappa(E)\big]}\,.
  \label{eq:Ldisp}
\end{equation}
The only physical-sheet singularity of $L$ is the threshold branch
point. The exit pole at $\kappa=-\gamma$ lies on the second sheet, and
even there the zero of $\Egam$ removes it, as the collapse to
Eq.~\eqref{eq:Ltri} below makes manifest: no anomalous threshold reaches
the physical region, and the entrance cut is the whole story.

Since $m_{N}\Egam = \gamma^{2}+\prel^{2}  = (\gamma+\kappa)(\gamma-\kappa)$, Eq.~\eqref{eq:Ldisp} is
\begin{equation}
  L(E) = \frac{A\,m_{N}\,\big[\gamma-\kappa(E)\big]}{4\pi},
  \label{eq:Ltri}
\end{equation}
linear in the branch variable: the deuteron pole and the vertex's photon
energy conspire to cancel. With $\mathcal{M}^{\mathrm{fact}}=-A$, Eq.~\eqref{eq:Tcap} gives
\begin{equation}
\begin{split}
  \mathcal{M}_{D}^{\mathrm{fact}}
  &= -A - \frac{4\pi/m_{N}}{\xi+\kappa}\,
    \frac{A\,m_{N}(\gamma-\kappa)}{4\pi}\\
  &= -A\,\frac{(\xi+\kappa)+(\gamma-\kappa)}{\xi+\kappa},
\end{split}
  \label{eq:assemble}
\end{equation}
and the $\kappa$ of the loop cancels the $\kappa$ of the denominator.
Substituting $A$ and stripping the common conversion factor between
$\mathcal{M}_{D}$ and the amplitude normalization of Eq.~\eqref{eq:AM1}, the
dressed factorized piece is
\begin{equation}
  \mathcal{N}^{\mathrm{fact}} = (1+\kappaV)\,\CSv\,\big[\gamma+\xi(\prel^{2})\big].
  \label{eq:Rimp}
\end{equation}

\subsubsection{Assembling ${\cal M}_D$}

The numerator of ${\cal M}_D$ assembles into:
\begin{equation}
\begin{split}
  \mathcal{N} &= \mathcal{N}^{\mathrm{fact}} + \mathcal{N}^{\mathrm{ct}}\\
  &= (1+\kappaV)\,\CSv\,\big[\gamma+\xi(\prel^{2})\big]
   + (\gamma^{2}+\prel^{2})\,\ell_{1},
\end{split}
  \label{eq:Nsum}
\end{equation}
This form is not unique: the UV part of the factorized piece feeds the contact term.  Mathematically, for any constant $c$,
\begin{equation}
\begin{split}
  (1+\kappaV)\,\CSv&\big[\gamma+\xi\big] + (\gamma^{2}+\prel^{2})\,\ell_{1}\\
  &= (1+\kappaV)\,\CSv\big[\gamma+\xi - c\,(\gamma^{2}+\prel^{2})\big]\\
  &\quad + (\gamma^{2}+\prel^{2})\,\big[\ell_{1}+c\,(1+\kappaV)\,\CSv\big],
\end{split}
  \label{eq:shuffle}
\end{equation}
representing the choice of scheme defining which finite terms are shuffled in or out of the renormalized $c_{\Mone}$ (inside $\ell_1$).
Equation~\eqref{eq:AM1} makes the
choice $c=\tfrac14(r_{0}+\rho_{d})$, which turns $\gamma+\xi$ into
the $W(\prel^{2})$ of Eq.~\eqref{eq:Wdef}, to echo the dibaryon literature~\cite{Ando:2005cz,Rupak:1999rk}.
The amplitude depends on $\rho_{d}$ only because of this choice of scheme.

The dressed matrix element is consistent with unitarity. Below the
inelastic threshold, cutting the nucleon pair relates the imaginary part
of $\mathcal{M}_{D}$ to the elastic amplitude on one side and
$\mathcal{M}_{D}$ itself on the other,
\begin{equation}
  \mathrm{Im}\,\mathcal{M}_{D}(s_{np})
  = \frac{\beta}{16\pi}\;4m_{N}^{2}\,
    T^{*}(s_{np})\,\mathcal{M}_{D}(s_{np}),
  \label{eq:watsoncut}
\end{equation}
with the same two-body phase space $\beta/16\pi$. Watson's
theorem~\cite{Watson:1954uc,Watson:1952ji} states that in a
time-reversal-invariant theory, a transition
amplitude feeding a strongly interacting two-body channel carries the
phase of the elastic scattering in that channel and nothing more:
$\arg\mathcal{M}_{D} = \delta_{^{1}S_{0}} \bmod \pi$.  On the cut the denominator
of Eq.~\eqref{eq:Rdef} carries exactly that phase,
$\xi - i\prel = |\xi+\kappa|\,e^{-i\delta_{^{1}S_{0}}}$, and
$\mathcal{N}$ is real: the assembled amplitude carries the elastic
phase as the theorem requires.

\subsubsection{Electric Channel}
\label{app:E1dress}

The same construction dresses the $\Eone$ amplitude. A single insertion of the
${}^{3}P$ amplitude of Eq.~\eqref{eq:Pwave} closes onto the tree amplitude
through Eq.~\eqref{eq:Tcap}, now in the $\ell=1$ channel: the cut carries the
centrifugal $\prel'^{3}$, and under it the production amplitude enters at
fixed photon energy, where the dipole matrix element behind
Eq.~\eqref{eq:AE1tree} carries a double deuteron pole,
$\mathcal{M}^{\Eone}(\prel')\propto\prel'/(\gamma^{2}+\prel'^{2})^{2}$
(on shell, one pole is absorbed into $\Egam$). The
discontinuity therefore grows like $\prel'^{3}$, and its dispersive
reconstruction costs subtractions absorbed into $c_{\Eone}$, exactly as the
boundary loop of Eq.~\eqref{eq:Lct} fed $c_{\Mone}$. The finite part that
survives is
\begin{equation}
  L \;\propto\; \prel\,
  \frac{\gamma\big(\gamma^{2}+3\prel^{2}\big)+2i\prel^{3}}
       {2\,\big(\gamma^{2}+\prel^{2}\big)^{2}}\,,
  \label{eq:LE1loop}
\end{equation}
and the double pole cancels against the tree amplitude in
$-T_{^{3}P_{J}}\,L/\mathcal{M}$.  Writing the elastic amplitude in terms of the
phase, $T_{^{3}P_{J}}=(4\pi/m_{N})\sin\delta_{^{3}P_{J}}e^{i\delta_{^{3}P_{J}}}/\prel^{3}$,
gives Eq.~\eqref{eq:dE1} directly.  The factor of two between the numerator of
Eq.~\eqref{eq:LE1loop} and Eq.~\eqref{eq:dE1} is carried by the $2$ in that
equation's denominator, which is the $8\pi$ rather than $4\pi$ normalization of the
loop; unitarity fixes it independently, through
$\operatorname{Im}G_{1}=m_{N}\prel^{3}/4\pi$, the $\ell=1$ value required by
$\operatorname{Im}\mathcal{M}_{D}=(m_{N}\prel^{3}/4\pi)\,T^{*}\mathcal{M}_{D}$.
Because one integral supplies both terms, their ratio is fixed and the overall
normalization drops out of Eq.~\eqref{eq:dE1} altogether.  The same result follows
from the coordinate-space effective-range calculation, where the dressing ratio is
$J_{G}/J_{F}$ with $J_{F,G}=\int_{0}^{\infty}\!dr\,r\,e^{-\gamma r}F_{1,G_{1}}(\prel r)$,
and from first-order perturbation theory in $a_{1}^{(J)}$.

The contact is a separate matter.  The dressing above was computed for the impulse
production amplitude, whose double deuteron pole controls the loop; the contact is a
point vertex, and the corresponding integral diverges and renormalizes the contact
itself.  Unitarity still forces the Watson phase, but the accompanying real
enhancement is the finite part of a subtracted loop, i.e.\ a higher-order contact.
We set it to zero, so that the contact is dressed by
$e^{i\delta_{^{3}P_{J}}}\cos\delta_{^{3}P_{J}}$ alone [Eq.~\eqref{eq:AE1}].
Refitting with the contact dressed instead exactly as the impulse term shifts
$c_{\Eone}$ by $+0.07$ while moving $\sigma(E)$ and the rate by less than
$0.04\%$; that scheme spread is carried in the truncation budget of
Sec.~\ref{sec:consequences}.  Written this
way the generalization to a coupled channel is immediate:
$\sin\delta\,e^{i\delta}=(S-1)/2i$, so it is the diagonal element of the $S$ matrix
that enters, and for ${}^{3}P_{2}$--${}^{3}F_{2}$ the bar convention gives
$S_{22}=\cos(2\varepsilon_{2})e^{2i\delta_{^{3}P_{2}}}$.  The mixing therefore acts
only through $\cos2\varepsilon_{2}$, quadratically in a parameter that the Nijmegen
analysis puts at $-0.175^{\circ}$ at the highest beam and $-3\times10^{-5}$ degrees
at $100\,\mathrm{keV}$; the ${}^{3}F_{2}$ component cannot reach the deuteron by
$\Eone$ at all.  Including it shifts $c_{\Eone}$ by $3\times10^{-4}$, and we drop it.
At leading
order in the phase the real part reduces to
$\delta_{\Eone}^{(J)}=(m_{N}\gamma/24\pi)(\gamma^{2}/3+\prel^{2})C_{p}^{(J)}$ with
$C_{p}^{(J)}=-36\pi\,a_{1}^{(J)}/m_{N}$, so that the constant of the averaged
treatment is
\begin{equation}
  C_{p}=\sum_{J}\frac{2J+1}{9}\,C_{p}^{(J)}
       =-\frac{4\pi}{m_{N}}\sum_{J}(2J+1)\,a_{1}^{(J)} .
  \label{eq:Cpsum}
\end{equation}
The weights are those of Eq.~\eqref{eq:sigma}, $w_{J}^{2}/8=(2J+1)/9$.  The
Nijmegen volumes $a_{1}^{(0,1,2)}=(-2.468,+1.529,-0.2844)\,\mathrm{fm}^{3}$~\cite{Stoks:1993tb,PavonValderrama:2004se}
give $C_{p}=-1.84\,\mathrm{fm}^{4}$, against the $-1.49\,\mathrm{fm}^{4}$ obtained
from the volumes implied by the 1994 potentials~\cite{Chen:1999bg,Rupak:1999rk}.
Because Eq.~\eqref{eq:Cpsum} cancels twelvefold, that $24\%$ spread traces almost
entirely to a $2.4\%$ difference in $a_{1}^{(^{3}P_{1})}$. The same cancellation is why
the volumes are quoted to four figures: rounding them to two changes $C_{p}$ by $3\%$.

\bibliographystyle{apsrev4-2}
\bibliography{refs}

\end{document}